\documentclass[AMA,STIX1COL]{WileyNJD-v2}
\usepackage{moreverb}
\usepackage{pdfpages}
\DeclareMathOperator{\Var}{Var}

\usepackage[colorinlistoftodos]{todonotes}

\newcommand\BibTeX{{\rmfamily B\kern-.05em \textsc{i\kern-.025em b}\kern-.08em
T\kern-.1667em\lower.7ex\hbox{E}\kern-.125emX}}

\articletype{Article Type}%

\received{<day> <Month>, <year>}
\revised{<day> <Month>, <year>}
\accepted{<day> <Month>, <year>}

\begin{document}

\title{Confidence Intervals for Prevalence Estimates from Complex Surveys with Imperfect Assays}

\author[1,2]{Damon M. Bayer}

\author[1]{Michael P. Fay*}

\author[3]{Barry I. Graubard}

\authormark{D. M. BAYER, M.P. FAY and B. I. GRAUBARD}

\address[1]{\orgdiv{Biostatistics Research Branch}, \orgname{National Institute of Allergy and Infectious Diseases}, \orgaddress{\state{Bethesda, Maryland}}}

\address[2]{\orgdiv{Department of Statistics}, \orgname{University of California, Irvine}, \orgaddress{\state{Irvine, California}}}

\address[3]{\orgdiv{Division of Cancer Epidemiology and Genetics}, \orgname{National Cancer Institute}, \orgaddress{\state{Rockville, Maryland}}}

\corres{*Michael P. Fay, Bethesda, MD 2089. \email{mfay@niaid.nih.gov}}


\abstract[Abstract]{We present several related methods for creating confidence intervals to assess disease prevalence in variety of survey sampling settings. These include simple random samples with imperfect tests, weighted sampling with perfect tests, and weighted sampling with imperfect tests, with the first two settings considered special cases of the third. Our methods use survey results and measurements of test sensitivity and specificity to construct melded confidence intervals. We demonstrate that our methods appear to guarantee coverage in simulated settings, while competing methods are shown to achieve much lower than nominal coverage. We apply our method to a seroprevalence survey of SARS-CoV-2 in undiagnosed adults in the United States between May and July 2020.}



\maketitle

\footnotetext{\textbf{Abbreviations:} ANA, anti-nuclear antibodies; APC, antigen-presenting cells; IRF, interferon regulatory factor}

\section{Introduction}

Estimating and quantifying uncertainty for disease prevalence is a standard task in epidemiology.
For rare events, these estimates are highly sensitive to misclassification \cite{hemenwaySelfDefense}, making adjustments for sensitivity and specificity critically important.
While estimating prevalence (or any event proportion in a population) in complex surveys and adjusting estimates for misclassification have been well studied separately, performing both of these tasks simultaneously remains relatively unexplored.
Recent overviews of methods for estimating prevalence in surveys without misclassification are provided by Dean and Pagano\cite{Dean:2015} and Franco, et al\cite{franco2019}.
For  simple random sample surveys with imperfect sensitivity and specificity, Lang and Reiczigel\cite{Lang:2014} proposed an approximate method that performed well in simulations. Recent work by DiCiccio, et al \cite{DiCi:2021} and Cai et al \cite{Cai:2020} study both valid (i.e., exact) and approximate methods. Their valid methods use test inversion and the adjustment of Berger and Boos \cite{Berg:1994}, while their approximation methods use the bootstrap with the test inversion approach. 
Fewer methods are available for for constructing frequentist confidence intervals for prevalence estimates from complex surveys while adjusting for sensitivity and specificity.
Kalish et al\cite{Kali:2021} developed one such method that is closely related to one of the methods presented here, but that method's properties were not studied.
Cai et al \cite{Cai:2020} (see also discussion in DiCiccio et al \cite{DiCi:2021}) modify their approximation approach to allow sample weights, but it assumes that the number of counts of events within the strata are large (see their Remark 4).  Thus, it would not apply to a weighted survey method where each individual their own weight. 
Another recent advancement is the method developed by Rosin et al \cite{rosin2021estimating} that 
 makes use of asymptotic normal approximations which reduce to the Wald interval when sensitivity and specificity are perfect.
This problem has also previously been addressed in Bayesian literature, recently by Gelman and Carpenter\cite{GelmanBayes}.

We work up to our ultimate goal in stages.
First, in Section~\ref{sec:srs-imperfect}, we propose confidence intervals for simple random samples where prevalence is assessed with an assay with imperfect sensitivity and/or specificity.
Next, in Section~\ref{sec:weight-perfect}, we present confidence intervals for weighted samples where prevalence is assessed with an assay without misclassification.
In Section~\ref{sec:weight-imperfect}, we combine these methods to create confidence intervals for weighted samples where prevalence is assessed with an assay with imperfect sensitivity and specificity.
Because the combined method reduces to one of the first two methods as a special case, we can think of the first two stages as testing the combined method in those cases.  Finally, in Section~\ref{sec:complex-surveys} we show how certain complex surveys may fit into the format for our new method.
Our new method is designed to guarantee coverage in all situations. 

In simulations we compare our method to established frequentist competitors.
and  show through simulations that it beats the best of those in each of the three stages with respect to guaranteeing coverage. 
We did not include in our simulations the new methods that have been developed in response to the COVID-19 pandemic and are not yet in print in peer reviewed journals \cite{Cai:2020,DiCi:2021,rosin2021estimating}.
The exact method of DiCiccio, et al \cite{DiCi:2021} would guarantee coverage, although applying it to a survey with a large number of strata would be ``computationally expensive'', and it has not been applied to surveys using post-stratification weighting. In contrast, our new method can very tractable in those situations.

\section{Confidence Interval Methods}

\subsection{Notation and Problem Set-up}
\label{sec-notation}

To introduce notation, consider first the stratified simple random sample.
Suppose we have a population partitioned into   \( K \) strata, with \( N_1, N_2, \ldots, N_K \) individuals in the  K strata of the population.
We sample \( n_1, n_2, \ldots, n_K \) individuals via a simple random sample from each of the \( K \)  strata to have an assay performed to determine who has a disease.
Let \( X_i \) be the number of positive results from an assay performed on the \( n_i \) individuals from stratum \( i \) and assume \( X_i \sim \operatorname{Binomial}(n_i, \theta_i) \), where \( \theta_i \) is the population frequency of positive results for assays performed on individuals from stratum \( i \).
Similarly, let \( X_i^* \) be the unobserved true number of people with the disease among the \( n_i \) individuals from stratum  \( i \) and assume \( X_i^* \sim \operatorname{Binomial}(n_i, \theta_i^*) \), where \( \theta_i^* \) is the population frequency of cases in stratum \( i \).
In the case of a perfect assay, \( \theta_i = \theta_i^* \).

Therefore, the population prevalence is 

\begin{equation}
    \beta^* = \frac{\sum_{i=1}^K N_i \theta_i^*}{\sum_{j=1}^K N_j} = \sum_{i=1}^K w_i \theta_i^*,
    \label{eq:pop-prev}
\end{equation}

and the apparent prevalence is 

\begin{equation}
    \beta = \frac{\sum_{i=1}^K N_i \theta_i}{\sum_{j=1}^K N_j} = \sum_{i=1}^K w_i \theta_i,
    \label{eq:app-prev}
\end{equation}

where \( w_i = N_i / \sum_{j=1}^K N_j \) and, therefore, \( \sum_{i=1}^K w_i = 1 \).
This set-up will approximately work for other complex survey samples, where we can estimate survey weights such that the complex survey sample may be treated as a multinomial sample with probabilities proportional to those weights  (see Section~\ref{sec:complex-surveys}).

We can relate $\theta_i$ and $\theta_i^*$ using the sensitivity ($\phi_p$)  and specificity (1-$\phi_n$) of the assay, 
where  $\phi_p$ and $\phi_n$  are the proportion of positive assays from a population of  positive controls (i.e., individuals known to have the disease) and 
negative controls (i.e., individuals known to be  without the disease), respectively. Then
$\theta_i = \phi_p \theta_i^* + \phi_n (1-\theta_i^*)$, or equivalently,  
\begin{eqnarray*}
\theta_i^* & = & \frac{ \theta_i - \phi_n }{\phi_p - \phi_n},
\end{eqnarray*}
and we have
\begin{align}
\begin{split}
  \beta^*   =&   \sum_{i=1}^K w_i \theta_i^* 
            =  \sum_{i=1}^K w_i \left( \frac{\theta_i - \phi_n}{\phi_p - \phi_n} \right) \\
            =&   \frac{\sum_{i=1}^K w_i \theta_i}{\phi_p - \phi_n} - \frac{\phi_n \sum_{i=1}^K w_i}{\phi_p - \phi_n} 
            =   \frac{\sum_{i=1}^K w_i \theta_i}{\phi_p - \phi_n} - \frac{\phi_n}{\phi_p - \phi_n}
            \label{eq:long-beta}
\end{split}
\end{align}

Suppose the assay is measured on \( m_n \) individuals known to not have the disease and on \( m_p \) individuals known to have the disease.
Let \( C_n \) and \( C_p \) be the number who test positive from the respective samples.
Assume that the negative and positive controls act like simple random samples from their respective populations.
Thus, \( C_n \sim \operatorname{Binomial}(m_n, \phi_n) \) where \( 1 - \phi_n \) is the specificity of the assay, and \( C_p \sim \operatorname{Binomial}(m_p, \phi_p) \), where  \( \phi_p \) is the sensitivity of the assay.
Let \( \hat{\theta}_i = \frac{X_i}{n_i} \), \( \hat{\phi}_n = \frac{C_n}{m_n} \), and \( \hat{\phi}_p = \frac{C_p}{m_p} \).
Then a plug-in estimator for \( \beta^* \) is 
\begin{equation}
    \hat{\beta}^* = \frac{\sum_{i=1}^K w_i \hat{\theta}_i}{\hat{\phi}_p - \hat{\phi}_n} - \frac{\hat{\phi}_n}{\hat{\phi}_p - \hat{\phi}_n}. \label{eq:betastarhat}
\end{equation}

This estimator serves as an important basis for developing confidence intervals in this work.
Section~\ref{sec:srs-imperfect} is concerned with confidence intervals for  \( \beta^* \) in the case where \( K = 1 \), \( \phi_n > 0 \), \( \phi_p < 1 \), i.e. estimating prevalence from a simple random sample with an imperfect assay.
Section~\ref{sec:weight-perfect} is concerned with confidence intervals for  \( \beta^* \) in the case where \( K > 1 \), \( \phi_n = 0 \), \( \phi_p = 1 \), i.e. estimating prevalence from a weighted sample with a perfect assay.
Section~\ref{sec:weight-imperfect} is concerned with confidence intervals for  \( \beta^* \) in the case where \( K > 1 \), \( \phi_n > 0 \), \( \phi_p < 1 \), i.e. estimating prevalence from a weighted sample with an imperfect assay.

\subsection{Estimating Prevalence from a Simple Random Sample with an Imperfect Assay}
\label{sec:srs-imperfect}

First, we consider the scenario where \( K = 1 \), \( \phi_n > 0 \), and \( \phi_p < 1 \).
We develop a confidence interval for the population prevalence, \( \beta^* \).
When \( K = 1 \), the estimand in Equation~\ref{eq:long-beta} becomes $\beta^* = (\theta_1 - \phi_n)/(\phi_p-\phi_n)$. We have $\phi_p > \phi_n$ for any useful assay, and since the sample is a mixture of individuals with and without the disease of interest, $\phi_p \geq \theta_1 \geq \phi_n$. The estimator of $\beta^*$ is 

\begin{equation}
\hat{\beta}^* \equiv 
g(\hat{\theta}_1, \hat{\phi}_n, \hat{\phi}_p)
\equiv 
\left\{ 
\begin{array}{ll}
1 & \mbox{ if $\hat{\phi}_n < \hat{\phi}_p < \hat{\theta}_1$ }  \\
\frac{\hat{\theta}_1 - \hat{\phi}_n}{\hat{\phi}_p - \hat{\phi}_n} & 
\mbox{ if $\hat{\phi}_p \geq \hat{\theta}_1 \geq \hat{\phi}_n$ } \\
0 & \mbox{ otherwise} 
\end{array}
\right.
\label{eq:srs-beta-est}
\end{equation}
where we define $0/0=0$.

To create a confidence interval for \( \hat{\beta}^* \), we use a generalization of the melding method \cite{FayP:2015}, which makes use of lower and upper confidence distributions on functions of independent estimators to account for variability in \( \hat{\theta}_1 \), \( \hat{\phi}_n \), and \( \hat{\phi}_p \). Confidence distributions are like frequentist posterior distributions\cite{Xie2013}.
The lower and upper confidence distributions are used with discrete responses to ensure the validity of the resulting inferences, and for the binomial case they are equivalent to the posterior distributions that result from using well-calibrated null preference priors \cite{Fay2021}.

Each estimated component in Equation~\ref{eq:srs-beta-est} is a binomial probability parameter.
For each of these, we use distributions associated with  the exact binomial confidence interval.
For a binomial experiment with \( x \) successes out of \( n \) trials, the lower confidence distribution is \( \operatorname{Beta}(x, n - x + 1) \) with associated random variable \( B^L \), and the upper confidence distribution  is \( \operatorname{Beta}(x + 1, n - x)\) with random variable \( B^U \), where for $a>0$ we let \( \operatorname{Beta}(0,a) \) and \( \operatorname{Beta}(a,0) \) be point masses at 0 and 1, respectively.
Let \( q(a, W) \) be the \( a \)th quantile of a random variable \( W \). Then the exact \( 1 - \alpha \)\% central confidence interval of Clopper-Pearson\cite{10.1093/biomet/26.4.404} for the binomial parameter is 
\begin{equation}
\left\{ q \left( \frac{\alpha}{2}, B^L \right), q \left( 1 - \frac{\alpha}{2}, B^U, \right) \right\}.
\label{eq:C-P}
\end{equation}

Fay et al \cite{FayP:2015} proposed a method for obtaining confidence intervals for functions of two parameters that are monotonic within the allowable range  for each parameter given the other is fixed. Here we generalize that to $\beta^*$, which is a function of 3 parameters. When $1 \geq \phi_p > \theta_1 > \phi_n \geq 0$ then $\beta^*$ is monotonically increasing in $\theta_1$, monotonically decreasing in $\phi_p$, and monotonically decreasing in $\phi_n$.
For an assessment of monotonicity in other scenarios see Appendix~\ref{monotonicity}.
Then the \( 1-\alpha \)\% confidence interval for \( \beta^* \) is 

\begin{equation}
    \left\{ q \left( \frac{\alpha}{2}, g \left\{ B_{\theta_1}^L, B_{\phi_n}^U, B_{\phi_p}^U \right\}   \right),  
            q \left( 1 - \frac{\alpha}{2},  g \left\{ B_{\theta_1}^U, B_{\phi_n}^L, B_{\phi_p}^L \right\}   \right) \right\}.
\label{eq:srs-conf-int}
\end{equation}
where $g(\cdot)$ is defined in equation~\ref{eq:srs-beta-est}.

The quantiles of these melded distributions are calculated by Monte Carlo sampling from each of the component distributions.
We compare this method to one described in Lang and Reiczigel \cite{Lang:2014} as implemented in prevSeSp function in \cite{asht}, which provides approximate confidence intervals for true prevalence when sensitivity and specificity are estimated from independent samples, as they are  in this section.
The Lang-Reiczigel interval is given by
\begin{equation}
\beta_1^{*\prime} + d\beta \pm q\left( 1 - \frac{\alpha}{2}, Z \right) \cdot \Var(\beta_1^{*\prime})^{1/2}    
\end{equation}

where \( q_Z \equiv q\left( 1 - \frac{\alpha}{2}, Z \right)\) and \( Z \sim N(0,1) \).

\begin{equation*}
    d\beta = 2 \cdot q_Z^2 \cdot\left\{ \beta_1^{*\prime} \cdot \frac{\phi_p^\prime (1 - \phi_p^\prime)}{m_p^\prime} - (1 - \beta_1^{*\prime}) \cdot \frac{(1 - \phi_n^\prime) \phi_n^\prime}{m_n^\prime}  \right\}
\end{equation*}

\begin{equation*}
    \Var(\beta_1^{*\prime}) = \frac{ \frac{\beta_1^{*\prime}(1 - \beta_1^{*\prime})}{n_1} + \left(\beta_1^{*\prime}\right)^2 \frac{\phi_p^\prime (1 - \phi_p^\prime)}{m_p} + \left(1 + \beta_1^{*\prime}\right)^2 \frac{(1 - \phi_n^\prime) \phi_n^\prime}{ m_n}}{(\phi_p^\prime - \phi_n^\prime)^2}
\end{equation*}

\begin{equation*}
    m_p^\prime = m_p +2
\end{equation*}

\begin{equation*}
    m_n^\prime = m_n + 2
\end{equation*}

\begin{equation*}
    \phi_p^\prime = \frac{m_p \cdot \hat{\phi}_p + 1}{m_p + 2}
\end{equation*}

\begin{equation*}
   1 - \phi_n^\prime = \frac{m_n \cdot (1 - \hat{\phi}_n) + 1}{m_n + 2} 
\end{equation*}

\begin{equation*}
   \beta_1^{*\prime} = \frac{\beta_1^\prime - \phi_n^\prime}{\phi_p^\prime - \phi_n^\prime} 
\end{equation*}
and 
\begin{equation*}
    \beta_1^\prime = \frac{n_1 \cdot \hat{\theta}_1 + q_Z^2 / 2}{n_1 + q_Z^2}.
\end{equation*}

\subsection{Estimating Prevalence from a Weighted Sample with a Perfect Assay}
\label{sec:weight-perfect}

Next, we present a confidence interval for the population prevalence, \( \beta^* \), in the scenario \( K > 1 \), \( \phi_n = 0 \), \( \phi_p = 1 \).
Our method is a straightforward adaptation of the gamma confidence interval presented in Fay and Feuer\cite{FayF:1997}, which was developed to create confidence intervals for a population rate which is assumed to be a weighted sum of Poisson rate parameters.
We note that for sufficiently large sample size \( n \) and small rate \( \lambda \), a \( \operatorname{Poisson}(n\lambda) \) distribution is approximately equal in distribution to a \( \operatorname{Binomial}(n, \lambda) \) distribution.
Under this Poisson assumption, we suggest the \( 100(1 - \alpha) \)\% gamma confidence interval for \( \beta^* \):

\begin{equation}
    \left( q\left( \frac{\alpha}{2}, G_{\beta^*}^L \right), q \left( 1 - \frac{\alpha}{2},  G_{\beta^*}^U \right) \right)
\end{equation}

where

\begin{equation*}
    G_{\beta^*}^L \sim \operatorname{Gamma}\left( \frac{y^2}{v}, \frac{v}{y} \right)
\end{equation*}

\begin{equation*}
    G_{\beta^*}^U \sim \operatorname{Gamma}\left( \frac{y^{*2}}{v^*}, \frac{v^*}{y^*} \right)
\end{equation*}

\begin{equation*}
    y = \sum_{i=1}^K \frac{w_i}{n_i} x_i
\end{equation*}

\begin{equation*}
    v = \sum_{i=1}^K \left( \frac{w_i}{n_i}\right)^2 x_i
\end{equation*}

\begin{equation*}
    y^* = y + \max\left(\frac{w_1}{n_1}, \ldots, \frac{w_K}{n_K} \right)
\end{equation*}

\begin{equation*}
    v^* = v + \left\{ \max\left(\frac{w_1}{n_1}, \ldots, \frac{w_K}{n_K} \right) \right\}^2.
\end{equation*}

We call this the wsPoison method, since it assumes a weighted sum of Poissons.
We compare the wsPoisson confidence interval to two methods presented in Dean and Pagano\cite{Dean:2015}, which were recommended for scenarios with low prevalence.
Dean and Pagano showed in simulations that the standard Wald interval had poor coverage with low prevalence (e.g., Fig. 1 of that paper showed 95\% confidence intervals with coverage of less than 85\% for prevalence values less than 2\%).
Since the confidence interval of Rosin, et al \cite{rosin2021estimating} reduces to the Wald interval with perfect assays, we will not include that method in the simulation comparisons.

The first recommended method of Dean and Pagano is an adaptation of the method of Agresti and Coull\cite{AgrestiCoull} for the survey setting.
The interval for \( \beta^* \) is given by:

\begin{equation}
    \tilde{p} \pm q_Z \sqrt{\tilde{p}(1 - \tilde{p}) / \tilde{n}}
\end{equation}

where 

\begin{equation*}
   \tilde{x} = \left( \sum_{i=1}^k w_i \hat{\theta}_i \right) n_{\text{eff}} + c 
\end{equation*}

\begin{equation*}
   \tilde{n} = n_{\text{eff}} + 2c 
\end{equation*}

\begin{equation*}
    \tilde{p} = \tilde{x} / \tilde{n}
\end{equation*}

\begin{equation*}
   c = q_Z^2/2
\end{equation*}

\begin{equation}
   n_{\text{eff}} = \frac{\left( \sum_{i=1}^k w_i \hat{\theta}_i \right) \left(1 - \sum_{i=1}^k w_i \hat{\theta}_i \right)}{\sum_{i=1}^k \frac{w_i^2}{n_i}\hat{\theta}_i} 
   \label{eq:neff}
\end{equation}
 
In the case where \( \sum_{i=1}^k \frac{w_i^2}{n_i}\hat{\theta}_i = 0 \), we instead let \( n_{\text{eff}} = \sum_{i=1}^k n_i \).

We also compare our suggested method to Dean and Pagano's modification of the method of Korn and Graubard\cite{Korn:1998,Dean:2015}.
This interval is given by 

\begin{equation}
    \left( q \left( \frac{\alpha}{2}, B^L_{KG} \right), q \left( 1 - \frac{\alpha}{2}, B^U_{KG} \right)  \right)
\end{equation}

where analogously to the Clopper-Pearson interval (see equation~\ref{eq:C-P}),
\begin{equation*}
    B^L_{KG} \sim \operatorname{Beta}\left(x_{\text{eff}},  n_{\text{eff}} -  x_{\text{eff}} + 1 \right)
\end{equation*}

\begin{equation*}
    B^U_{KG} \sim \operatorname{Beta}\left(x_{\text{eff}} + 1, n_{\text{eff}} - x_{\text{eff}} \right)
\end{equation*}

with \( x_{\text{eff}} = n_{\text{eff}} \sum_{i=1}^k w_i \hat{\theta}_i \), and \( n_{\text{eff}} \) defined in Equation~\ref{eq:neff}.
Although Dean and Pagano \cite{Dean:2015} expressed this in terms of the F distributions, the beta distribution representation is equivalent.

\subsection{Estimating Prevalence from a Weighted Sample with an Imperfect Assay}
\label{sec:weight-imperfect}

Lastly, we develop a confidence interval for the population prevalence, \( \beta^* \), in the case where \( K > 1 \), \( \phi_n > 0 \), \( \phi_p < 1 \).

The two methods we discuss are closely related to each other and the methods discussed in Sections~\ref{sec:srs-imperfect} and \ref{sec:weight-perfect}.
As in Section~\ref{sec:srs-imperfect}, we use the melding method \cite{FayP:2015} to create \( 1 - \alpha \)\% confidence interval very similar to Equation~\ref{eq:srs-conf-int}.
The confidence distributions for \( \phi_p \) and \( \phi_n \) are the same Beta distributions as in Section~\ref{sec:srs-imperfect}.
The two methods differ in their confidence distributions for the apparent prevalence \( \beta \).

In the first case, we use the adaptation of the gamma confidence interval \cite{FayF:1997} presented in Section~\ref{sec:weight-perfect} to derive the \( 1 - \alpha \)\% confidence interval for \( \beta^* \):

\begin{equation}
    \left\{ q \left( \frac{\alpha}{2}, 
    g \left[ G^L_{\beta^*}, B^U_{\phi_n}, B^U_{\phi_p} \right]
    \right),  q \left( 1 - \frac{\alpha}{2}, 
       g \left[G^U_{\beta^*}, B^L_{\phi_n}, B^L_{\phi_p} \right] \right)
       \right\}
\end{equation}

where  \( G_{\beta^*}^L \) and \( G_{\beta^*}^U \) are defined in Section~\ref{sec:weight-perfect}.
We refer to this method as the WprevSeSp Poisson - weighted prevalence with sensitivity and specificity, where the prevalence confidence distribution is based on the weighted sum of Poissons.

The alternative method is very similar to that used in Kalish, et al \cite{Kali:2021}.
We use the  modification of \cite{Korn:1998} presented in \cite{Dean:2015}, as in Section~\ref{sec:weight-perfect}, to derive the \( 100(1 - \alpha) \)\% confidence interval for \( \beta^* \):

\begin{equation}
    \left\{ q \left( \frac{\alpha}{2}, g\left[B_{KG}^L, B_{\phi_n}^U, B_{\phi_p}^U\right]  \right),  q \left( 1 - \frac{\alpha}{2}, g\left[B_{KG}^U, B_{\phi_n}^L, B_{\phi_p}^L\right]  \right) \right\}.
\end{equation}

where \( B_{KG}^L \) and \( B_{KG}^U \) are as defined in Section~\ref{sec:weight-perfect}. Although equivalent, this expression looks different than in Kalish, et al\cite{Kali:2021} because they used a parameter for specificity, rather than $\phi_n$, which is 1 minus specificity. 
We refer to this as WprevSeSp Binomial - weighted prevalence with sensitivity and specificity, where the prevalence confidence distribution is based on a binomial variance assumption.

\subsection{Applications to More Complex Surveys}
\label{sec:complex-surveys}

\subsubsection{When Can We Use These Methods}

In Section~\ref{sec-notation}, we derived methods assuming that the apparent prevalence was a weighted sum of binomial random variables,
$\beta = \sum_{i=1}^{K} w_i \frac{X_i}{n_i}$, where $X_i \sim \textrm{Binomial}(n_i,\theta_i)$. We used the fact that for small $\theta_i$ and large $n_i$
the binomial can be approximated by the Poisson, giving $X_i \stackrel{\cdot}{\sim} \textrm{Poisson}( n_i \theta_i)$.
Thus, whenever we can model a complex survey estimator of apparent prevalence as a weighted sum of Poisson variates, then we can apply the methods of this paper.

In the upcoming Section~\ref{sec-MultPoisson}, we give a detailed review relating the multinomial sampling model  to a weighted sum of Poisson variates model.
The multinomial sampling model treats the survey sample as if it is a sampling with replacement from the entire population of $N$ individuals, where each of the $N$
individuals has a probability of $P_j$ of being sampled for each of the $n$ samples from the survey, with $\sum_{j=1}^{N} P_j = 1$. Under this model the number of times each of the $N$ individuals
is included in the sample is a multinomial with parameters $n$ and $[P_1,\ldots, P_N]$.
The multinomial model describes sampling with replacement, but it is nevertheless used to approximate a sampling design where the $j$th  individual is sampled {\it without} replacement with probability $P_j$, even though under that design (unlike the multinomial model) no individual is included in the sample more than once. The multinomial model is a common approximation for other complex survey designs; \citep[see e.g., ][p. 14]{Korn:1999}.
For example, in the  Kalish, et al\cite{Kali:2021} analysis of Section~\ref{sec-Application} each individual in the sample is assigned a pseudo-weight approximating one over their sampling probability from
a multinomial model. The actual sample was not a probability sample. In fact, it was a quota sample from a very large pool of self-selected volunteers, and the pseudo-weights were calculated using a different large survey that was a probability weighted survey. The pseudo-weights were calculated such that if they were analyzed under the multinomial model, they would adjust for selection bias due to self-selection of the volunteers and the imperfection of the quota sampling.

\subsubsection{Multinomial Sampling Model}
\label{sec-MultPoisson}

Let $Y_1,\ldots, Y_N$ be the binary indicators of event in the $N$ individuals  in the population of interest,
so the prevalence is $\beta = N^{-1} \sum_{j=1}^{N} Y_j$. There are many ways to design a complex survey sample, and it is often useful to analyze them as if
individuals were sampled with replacement with the sampling probability of the $j$th  individual equal to $P_j$, with $\sum_{j=1}^{N} P_j = 1$
In other words, we treat the sample as if it was $n$ independent multinomial samples each with one trial and selection probability vector  $[P_1,\ldots, P_N]$.
Let $I_{ij}=1$ if the $i$th draw for the sample is individual $j$ in the population, and $0$ otherwise. Then let $y_i = Y_j$ and $p_i=P_j$ when $I_{ij}=1$.
Here, following the tradition in the survey literature, we use capital letters for the population of interest (e.g., N,Y,P), and lower case letters for the sample (e.g., n,y,p).
In this notation, both $y_i$ and $Y_j$ are fixed, and only the variables representing the sampling (i.e., the $I_{ij}$ variables) are random.
Under this independent multinomial model, since $E(I_{ij})=P_j$, an unbiased estimator of $\beta$ is
\begin{eqnarray}
\hat{\beta} & = & \frac{1}{n} \sum_{i=1}^{n} \frac{ y_i}{N p_i} = \frac{1}{n} \sum_{i=1}^{n} \sum_{j=1}^{N}  \frac{ I_{ij}  Y_j}{N P_j}
\label{eq:betahatMultinomial}
\end{eqnarray}
and an unbiased estimator of $\textrm{var}(\hat{\beta})$ under the multinomial model is
\begin{eqnarray}
\widehat{\textrm{var}}_{M}(\hat{\beta}) & = & \frac{1}{n (n-1)} \sum_{i=1}^{n} \left( \frac{y_i}{Np_i} - \hat{\beta}  \right)^2
\label{eq:varMbetahat}
\end{eqnarray}
(see Korn and Graubard \cite{Korn:1999} Problem 2.2-10). We can write $\hat{\beta}$ as a weighted sum. Traditional survey weighting defines the weights so that
the weight for the $i$th sampled individual can be interpreted as the number of individuals in the population that the $i$th sampled individual represents.
Following that tradition, let   $w_i^{(trad)}= 1/(np_i)$ and $W_j^{trad)}=1/(nP_j)$, then the expected sum of the sampled weights is $N$,
\begin{eqnarray*}
E \left( \sum_{i=1}^{n} w_i^{(trad)} \right)  & = & E \left(  \sum_{i=1}^{n} \sum_{j=1}^{N} I_{ij} W_j^{(trad)} \right) =    \sum_{i=1}^{n} \sum_{j=1}^{N} \frac{ E(I_{ij})}{ n P_j} =   \sum_{i=1}^{n} \sum_{j=1}^{N} \frac{1}{n} = N.
\end{eqnarray*}
Sometimes the weights are scaled after selection so that the scaled weights are $w_i^{(strad)} = \frac{ N w_i^{(trad)} }{\sum_{i=1}^{n} w_i^{(trad)}}$ and are forced to sum to $N$.
For example, in Kalish et al \cite{Kali:2021} rescaling (sometimes called post-stratification) was done in a more complicated manner to ensure that the weights summed to the US census population within age group, sex, race, ethnicity and region.

For this paper we define the weights differently, because we want to model our estimator as a weighted sum of Poisson random variables.
Thus, we use $w_i = 1/(nNp_i)$ and $W_j= 1/(nNP_j)$ so that the sums have expectation $1$.
In the complex survey case,
we start with the independent multinomial model as in equation~\ref{eq:betahatMultinomial}, then we use the relationship between the multinomial and Poisson distributions.
Using the ``multinomial-Poisson transformation'', the maximum
likelihood estimates (MLE) for a multinomial random variable are equivalent to the MLEs for independent Poisson random variables, and
the variances are asymptotically equivalent (see Baker\cite{Baker:1994}).
Even though we model $\hat{\beta}$ using multinomial random variables where there are many missing values (which occurs in our situation whenever $I_{ij}=0$),
that multinomial-Poisson relationship holds even when there are missing variables (see Baker\cite{Baker:1994}, Section 3).
For both the Poisson and multinomial models, $E(I_{ij}) = P_j$, and $\hat{\beta}$ is unbiased under either model.
For the Poisson model, all the $I_{ij}$ are independent and each mean equals its variance, so that the variance of $\hat{\beta}$ under this model is
\begin{eqnarray*}
\textrm{var}_P \left(\hat{\beta} \right) & = & \textrm{var}_P \left(  \frac{1}{n}  \sum_{i=1}^{n} \sum_{j=1}^{N}  \frac{ I_{ij}  Y_j}{N P_j} \right)  \\
& = &    \sum_{i=1}^{n} \sum_{j=1}^{N}  \frac{ \textrm{var}_P( I_{ij})  Y_j^2}{n^2 N^2 P_j^2}  \\
& = &    \sum_{i=1}^{n} \sum_{j=1}^{N}  \frac{   Y_j}{n^2 N^2 P_j}.
\end{eqnarray*}
We estimate $\textrm{var}_P \left(\hat{\beta} \right)$ by multiplying each term in the sum by $I_{ij}/P_j$, which has an expectation of $1$ and eliminates terms of non-selected individuals, giving
\begin{eqnarray}
\widehat{\textrm{var}}_P \left(\hat{\beta} \right)
& = &   \sum_{i=1}^{n} \sum_{j=1}^{N}  \frac{ I_{ij}  Y_j}{n^2 N^2 P_j^2}  =  \sum_{i=1}^{n}  \frac{  y_i}{n^2 N^2 p_i^2} \label{eq:hatvarbetahat1}
\end{eqnarray}
Under the Poisson model $\widehat{\textrm{var}}_P \left(\hat{\beta} \right)$ is an unbiased estimator of $\textrm{var}_P \left(\hat{\beta} \right)$.

\section{Simulations}

\subsection{Estimating Prevalence from a Simple Random Sample with an Imperfect Assay}

\begin{figure}
    \centering
    \includegraphics[width=0.8\textwidth]{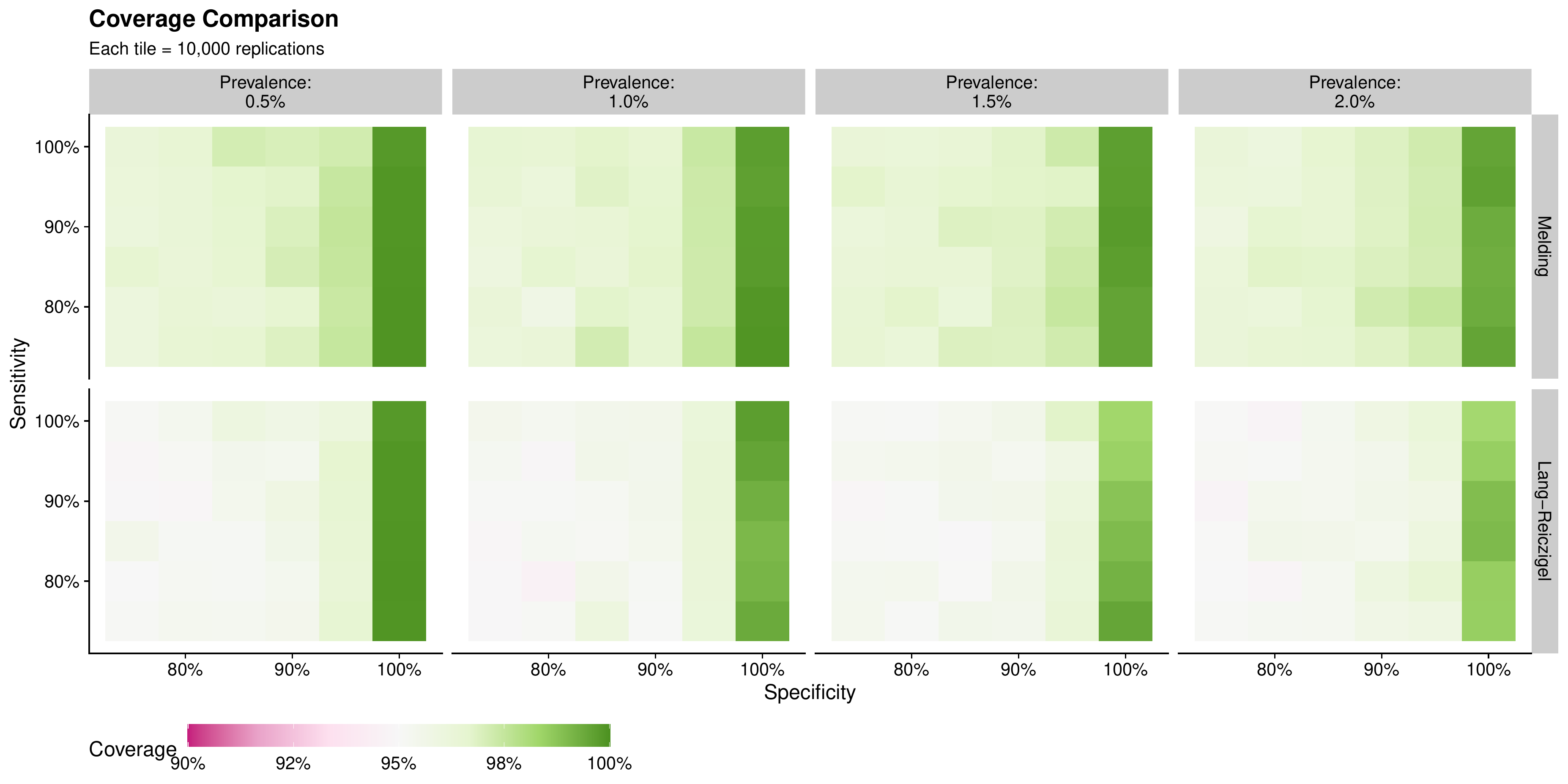}
    \caption{Coverage properties of 95\% confidence intervals for our new method (Melding) and the Lang-Reiczigel method in a variety of settings, each simulated 10,000 times.}
    \label{fig:coverage_comparison_plot}
\end{figure}

\begin{figure}
    \centering
    \includegraphics[width=0.8\textwidth]{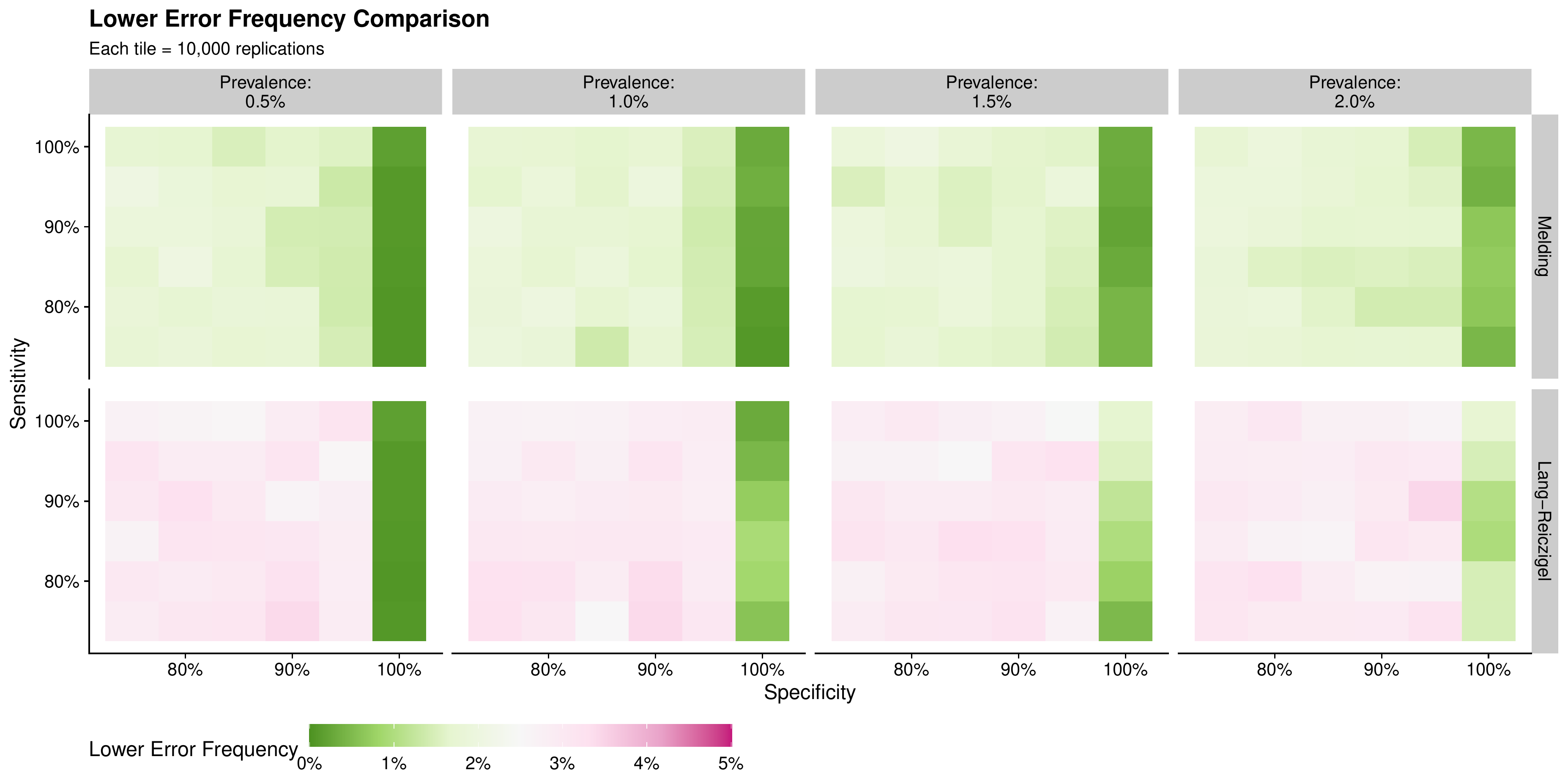}
    \caption{Lower error properties of 95\% confidence intervals for our new method (Melding) and the Lang-Reiczigel method in a variety of settings, each simulated 10,000 times.}
    \label{fig:lower_error_frequency_comparison_plot}
\end{figure}

\begin{figure}
    \centering
    \includegraphics[width=0.8\textwidth]{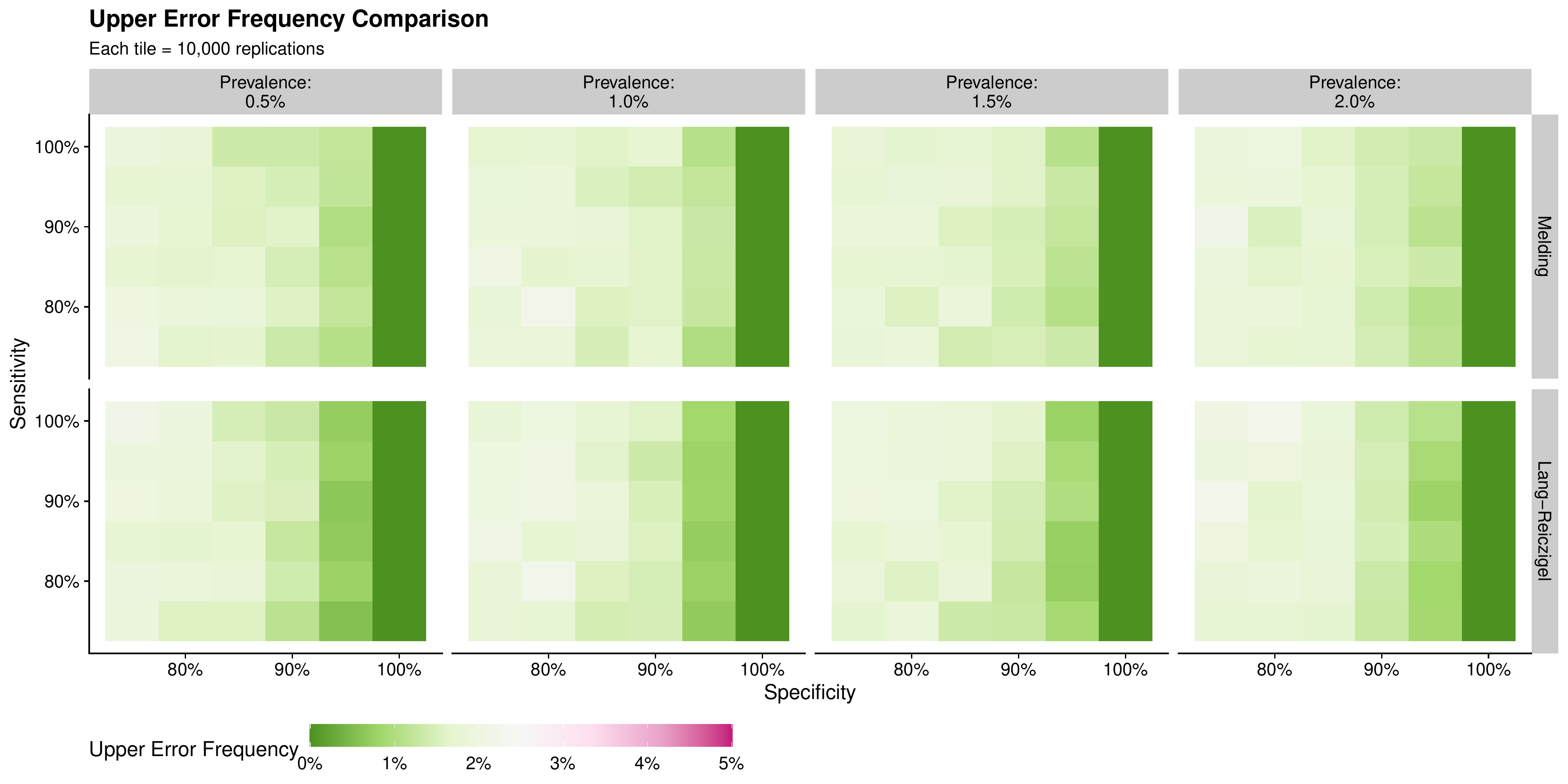}
    \caption{Upper error properties of 95\% confidence intervals for our new method (Melding) and the Lang-Reiczigel method in a variety of settings, each simulated 10,000 times.}
    \label{fig:upper_error_frequency_comparison_plot}
\end{figure}

We assess and compare our new method (Melding, i.e., equation~\ref{eq:srs-conf-int}) to that of Lang and Reiczigel (LR) in a variety of simulated settings.
In each simulation, 100 subjects are tested to estimate prevalence, 60 are tested to estimate sensitivity, and 300 are tested to estimate specificity.
Several combinations of prevalences (0.5\%--2\%), sensitivities (75\%--100\%) and specificities (75\%--100\%) are assessed.
Each simulated scenario is replicated 10,000 times.
Figure~\ref{fig:coverage_comparison_plot} compares the two methods based on coverage, while Figures~\ref{fig:lower_error_frequency_comparison_plot} and \ref{fig:upper_error_frequency_comparison_plot} present the lower and upper error frequencies for these scenarios, respectively.

Figure~\ref{fig:coverage_comparison_plot} shows that, when specificity is less than perfect, both methods achieve approximately nominal coverage, with the melding method being somewhat more conservative.
When specificity is 100\%, both methods are conservative.
Figure~\ref{fig:upper_error_frequency_comparison_plot} shows that both methods make upper errors with roughly the same frequency.
Figure~\ref{fig:lower_error_frequency_comparison_plot} demonstrates that while the melding procedure bounds the lower error frequency below 2.5\%, the Lang-Reiczigel method generally has lower error above 2.5\%, which is undesirable for applications for which there is a need to bound the lower errors.

\subsection{Estimating Prevalence from a Weighted Sample with a Perfect Assay}
\label{sim-perfect}
We compare the wsPoisson method to the more traditional Dean-Pagano modification of the Agresti-Coull (DPAC) method and the Korn-Graubard (KG) method for survey proportions in a variety of settings.
Our simulations examine varying levels of disease prevalence (0.5\% or 5\%), different types  survey designs (50 sampling strata with  200 subjects each or 8000 individuals, each with their own weight), distributions of weights among the sampling strata or individuals (coefficient of variation from approximately 0\% to nearly 600\%), and the number and weights of sampling strata with non-zero prevalence.
For each combination of prevalence \( p \), and group type, up to 500 sets of weights are simulated.
These 500 sets of weights are designed to span a range of coefficients of variation.
For a target coefficient of variation, \( v \), \( n \) weights (\(w_i\)) are simulated by generating \( n \) samples from a \( \text{Beta} \left(\frac{1}{v^2} - \frac{1}{nv^2} - \frac{1}{n}, \frac{n-1}{v^2} - \frac{n-1}{nv^2} - \frac{n-1}{n} \right) \) distribution and normalizing so that \( \sum_{i=1}^n w_i = 1 \).
This assures that the coefficient of variation among these weights is approximately \( v \).
Then, certain weights are chosen to have non-zero prevalence (5\%, 25\%, or 75\% distributed either among the highest weights, lowest weights, or distributed uniformly).
These weights with non-zero prevalence are given a prevalence such that \( \sum_{i=1}^n w_i \theta_i = p \).
For each simulated set of parameters and weights, 10,000 data sets are simulated and assessed. 

\begin{figure}
\centering
\includegraphics[width=0.8\textwidth]{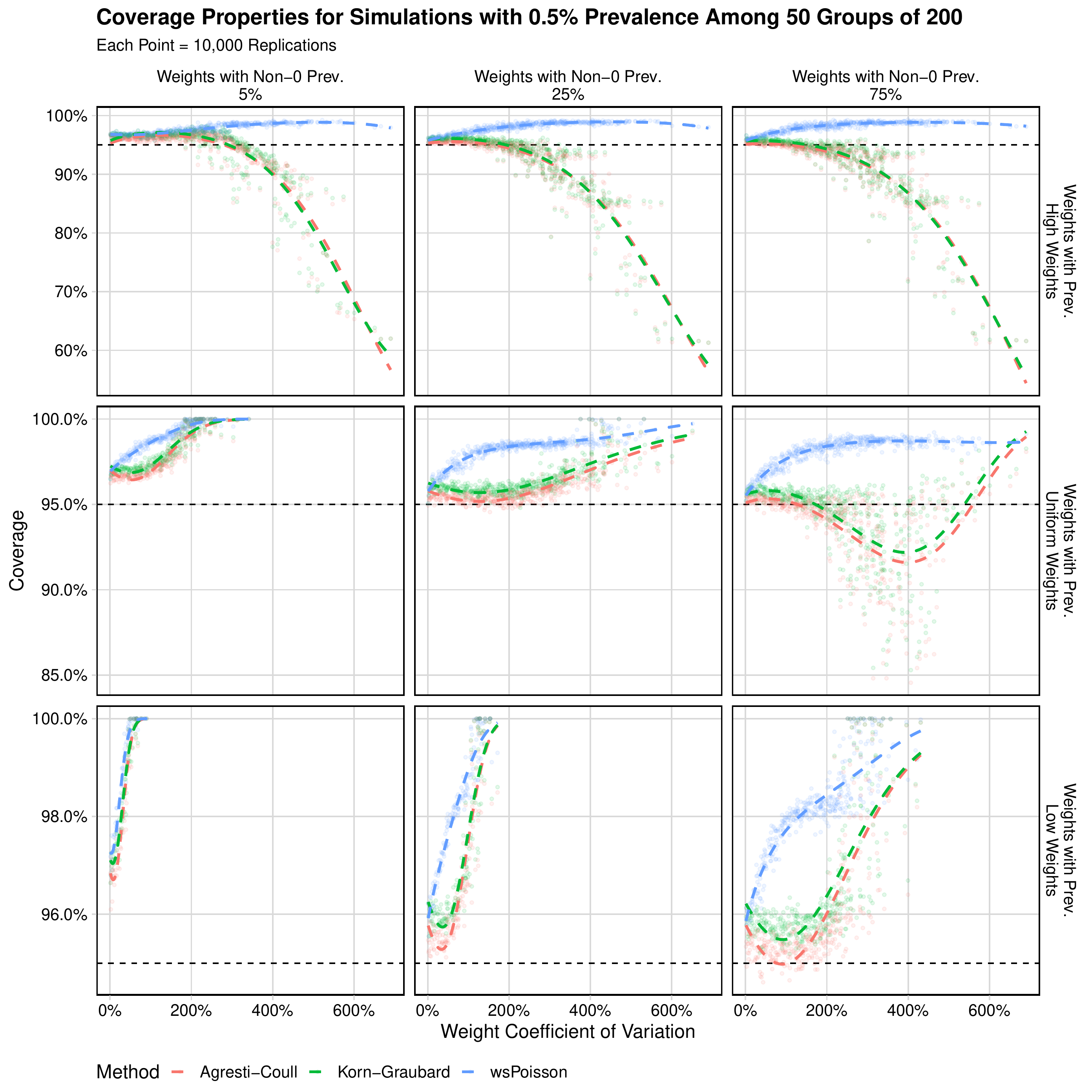}
\caption{Coverage properties for the wsPoisson model and two standard methods, the Dean-Pagano modification of the Agresti-Coull method and of the Korn-Graubard method.
Each point represents 10,000 simulations of datasets from a population with 0.5\% Prevalence where 50 groups of 200 people are sampled.
The horizontal dashed line indicates the nominal coverage, 95\%.
Colored dashed lines are estimates from a logistic regression model using quadratic splines.}
\label{fig:perfect_coverage_50_groups_0_005_prev}
\end{figure}

\begin{figure}
\centering
\includegraphics[width=0.8\textwidth]{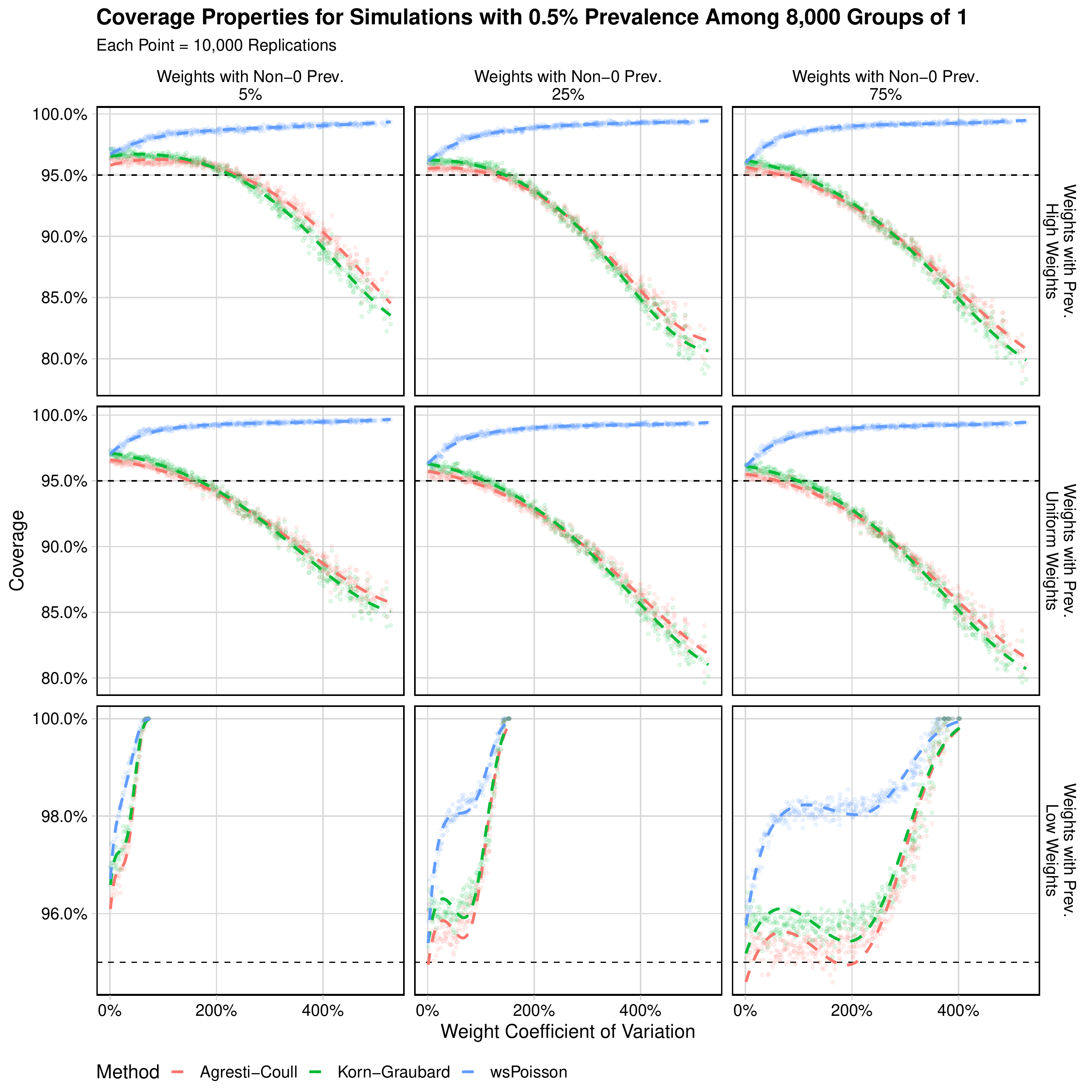}
\caption{Coverage properties for the wsPoisson model and two standard methods, the Dean-Pagano modification of the Agresti-Coull method and of the Korn-Graubard method.
Each point represents 10,000 simulations of datasets from a population with 0.5\% Prevalence where 8000 individuals are sampled.
The horizontal dashed line indicates the nominal coverage, 95\%.
Colored dashed lines are estimates from a logistic regression model using quadratic splines.}
\label{fig:perfect_coverage_8000_groups_0_005_prev}
\end{figure}

\begin{figure}
\centering
\includegraphics[width=0.8\textwidth]{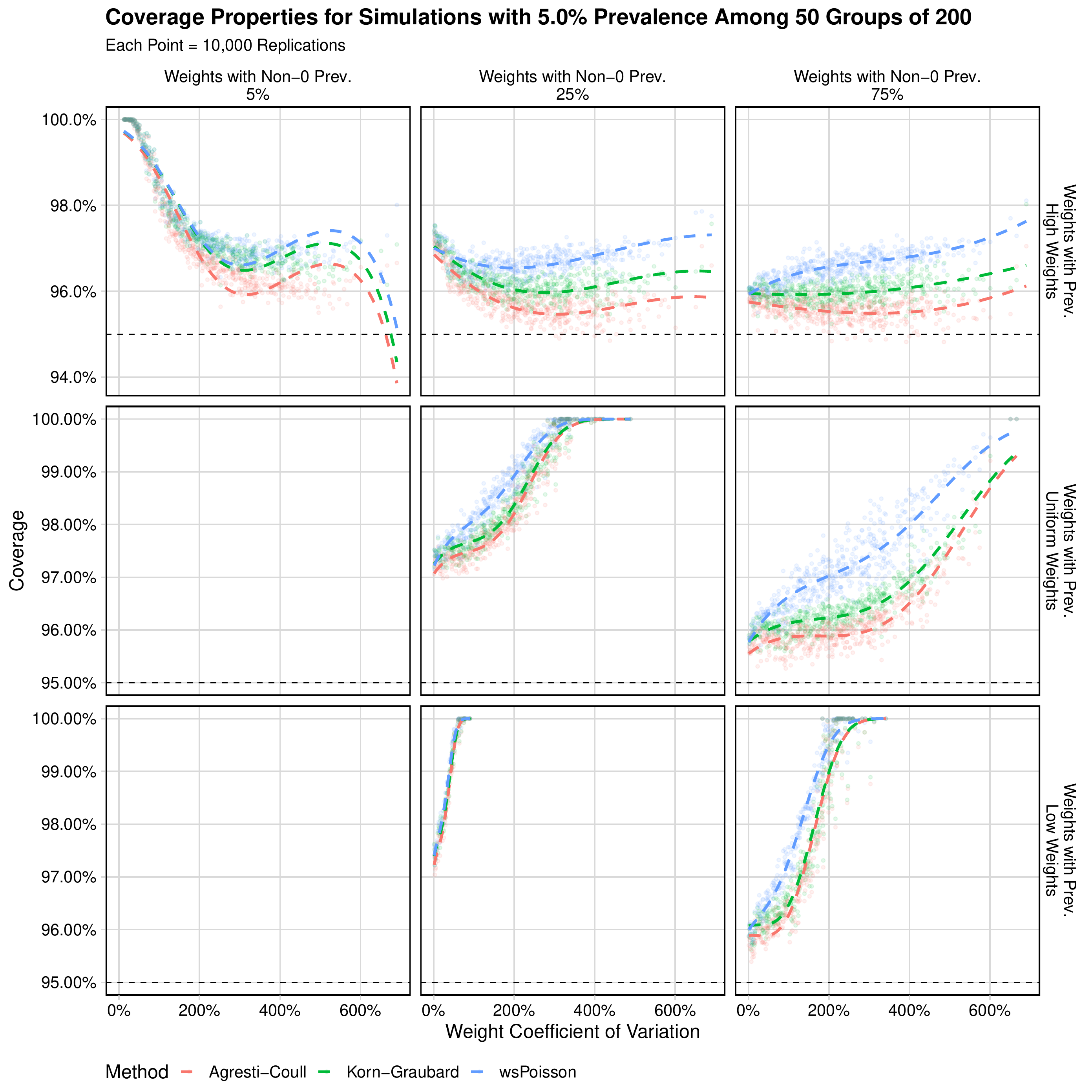}
\caption{Coverage properties for the wsPoisson model and two standard methods, the Dean-Pagano modification of the Agresti-Coull method and of the Korn-Graubard method.
Each point represents 10,000 simulations of datasets from a population with 5\% Prevalence where 50 groups of 200 people are sampled.
The horizontal dashed line indicates the nominal coverage, 95\%.
Colored dashed lines are estimates from a logistic regression model using quadratic splines.}
\label{fig:perfect_coverage_50_groups_0_05_prev}
\end{figure}

\begin{figure}
\centering
\includegraphics[width=0.8\textwidth]{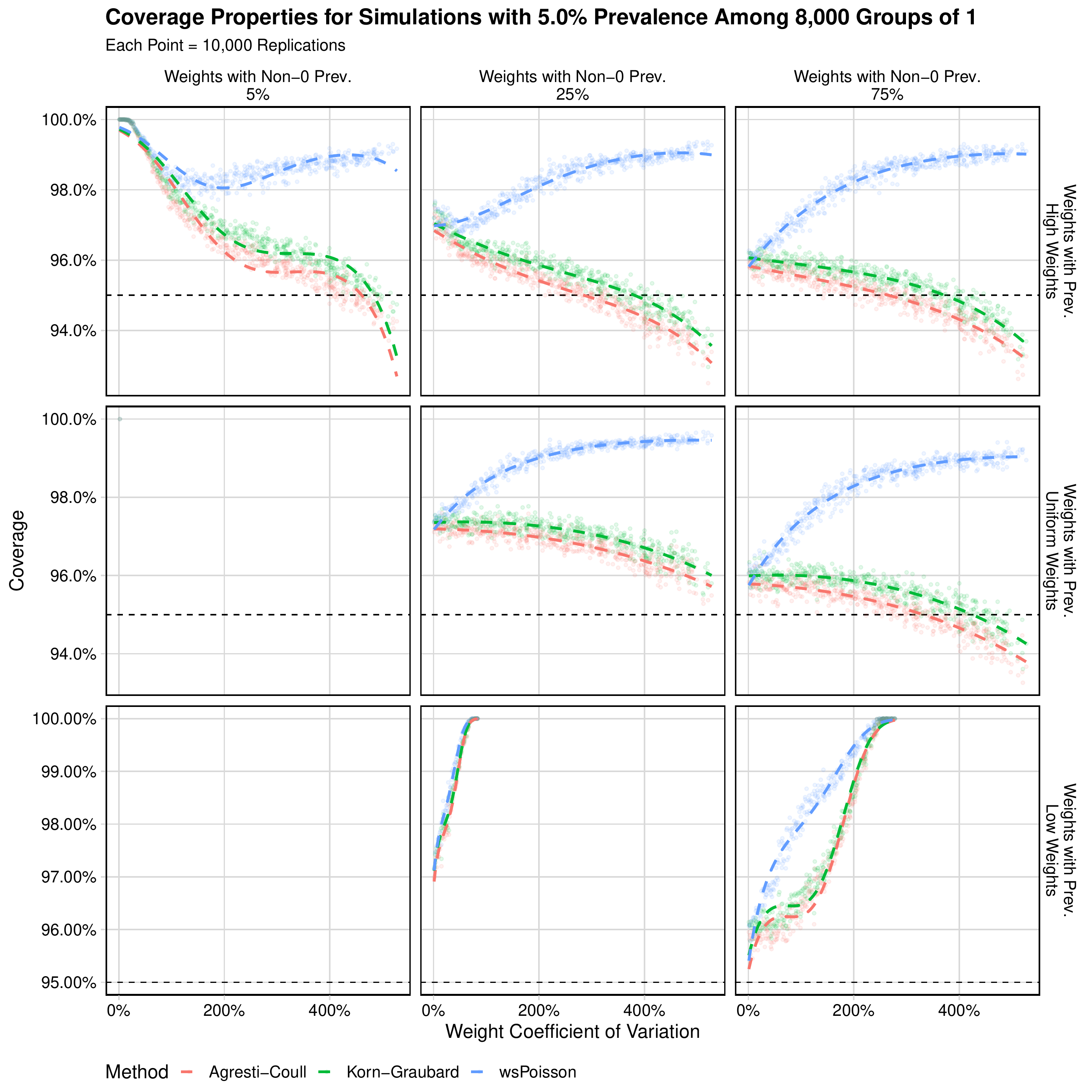}
\caption{Coverage properties for the wsPoisson model and two standard methods, the Dean-Pagano modification of Agresti-Coull and Korn-Graubard
Each point represents 10,000 simulations of datasets from a population with 5\% Prevalence where 8000 individuals are sampled.
The horizontal dashed line indicates the nominal coverage, 95\%.
Colored dashed lines are estimates from a logistic regression model using quadratic splines.}
\label{fig:perfect_coverage_8000_groups_0_05_prev}
\end{figure}

The coverage properties for these simulations are presented in Figures~\ref{fig:perfect_coverage_50_groups_0_005_prev}--\ref{fig:perfect_coverage_8000_groups_0_05_prev}.
Additional properties for these simulations are presented in Figures~\ref{fig:perfect_lower_error_frequency_50_groups_0_005_prev}--\ref{fig:perfect_confidence_interval_width_8000_groups_0_05_prev}.

From Figures~\ref{fig:perfect_coverage_50_groups_0_005_prev}--\ref{fig:perfect_coverage_8000_groups_0_05_prev}, we note that the two competitor methods generally exhibit lower coverage as the coefficient of variation among the weights increases.
In Figure~\ref{fig:perfect_coverage_50_groups_0_005_prev}, this coverage falls below 60\% when the prevalence very low and is concentrated among the highest weights, and the coefficient of variation among the weights exceeds 4.
Uniform distribution of prevalence among the weights, increased overall prevalence, and larger sample sizes among fewer groups all appear to lessen the severity of this problem.
In contrast, the wsPoisson method appears to guarantee coverage in all scenarios. The wsPoisson method tends to become more conservative when the coefficient of variation among the weights increases, when the other methods can have problems guaranteeing coverage. 
In all cases, the wsPoisson method is more conservative than the competitor methods.
This is similar to the behavior observed in Fay and Feuer\cite{FayF:1997}, where, in simulations, the overall error rate for the gamma intervals decreased as the variance of the weights increased.
Because our methods appear to be very conservative, with coverage near 100\% in some cases, we present the widths of the confidence intervals in Figures~\ref{fig:perfect_confidence_interval_width_50_groups_0_005_prev}--\ref{fig:perfect_confidence_interval_width_8000_groups_0_05_prev}.
In scenarios where coefficient of variation among the survey weights is high, the wsPoisson intervals are often two or three times wider than intervals produced by competing methods.

\subsection{Estimating Prevalence from a Weighted Sample with an Imperfect Assay}

We compare properties our melded confidence interval WprevSeSp Poisson, to another melded confidence interval method WprevSeSp Binomial, and one method, wsPoisson, which does not account for the imperfect assay.
The methods are assessed in a several simulated scenarios with varying levels of disease prevalence (0.5\% or 5\%), types of groups surveyed (50 groups 200 subjects or 8000 groups of 1 subject), distributions of weights among the groups (coefficient of variation from approximately 0\% to nearly 6\%), the number of groups with non-zero prevalence, and the specificity of the assay (80\% - 100\%).
In each scenario, the assay has 95\% sensitivity.
Each scenario creates up to 500 new sets of weights and $\theta_i$ parameters (as in Section~\ref{sim-perfect}), and each of those is simulated 10,000 times, with new prevalence, sensitivity, and specificity surveys generated and 95\% confidence intervals are created.
Modelled after the study of Kalish, et al \cite{Kali:2021}, the simulated sensitivity is assessed based on 60 tests, while specificity is based on 300 tests.

\begin{figure}
\centering
\includegraphics[width=0.8\textwidth]{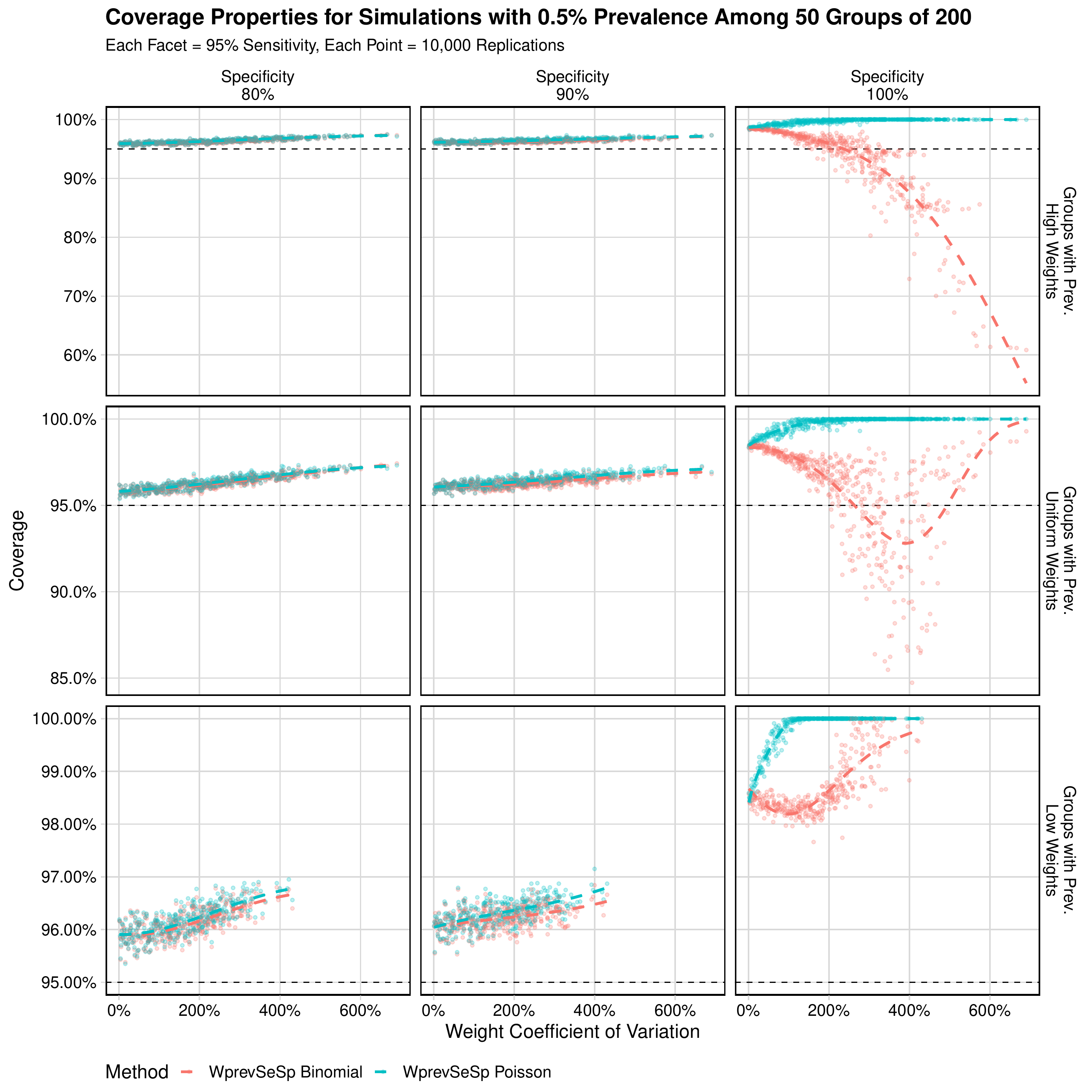}
\caption{Coverage properties for the confidence interval procedures, WprevSeSp Binomial and WprevSeSp Poisson.
Each point represents 10,000 simulations of datasets from a population with 0.5\% Prevalence where 50 groups of 200 people are sampled.
Each datasets also includes simulated results of tests to evaluate the sensitivity and specificity of the assay performed on 60 and 300 individuals, respectively.
The horizontal dashed line indicates the nominal coverage, 95\%.
Colored dashed lines are estimates from a logistic regression model using quadratic splines.}
\label{fig:imperfect_coverage_50_groups_0_005_prev}
\end{figure}

\begin{figure}
\centering
\includegraphics[width=0.8\textwidth]{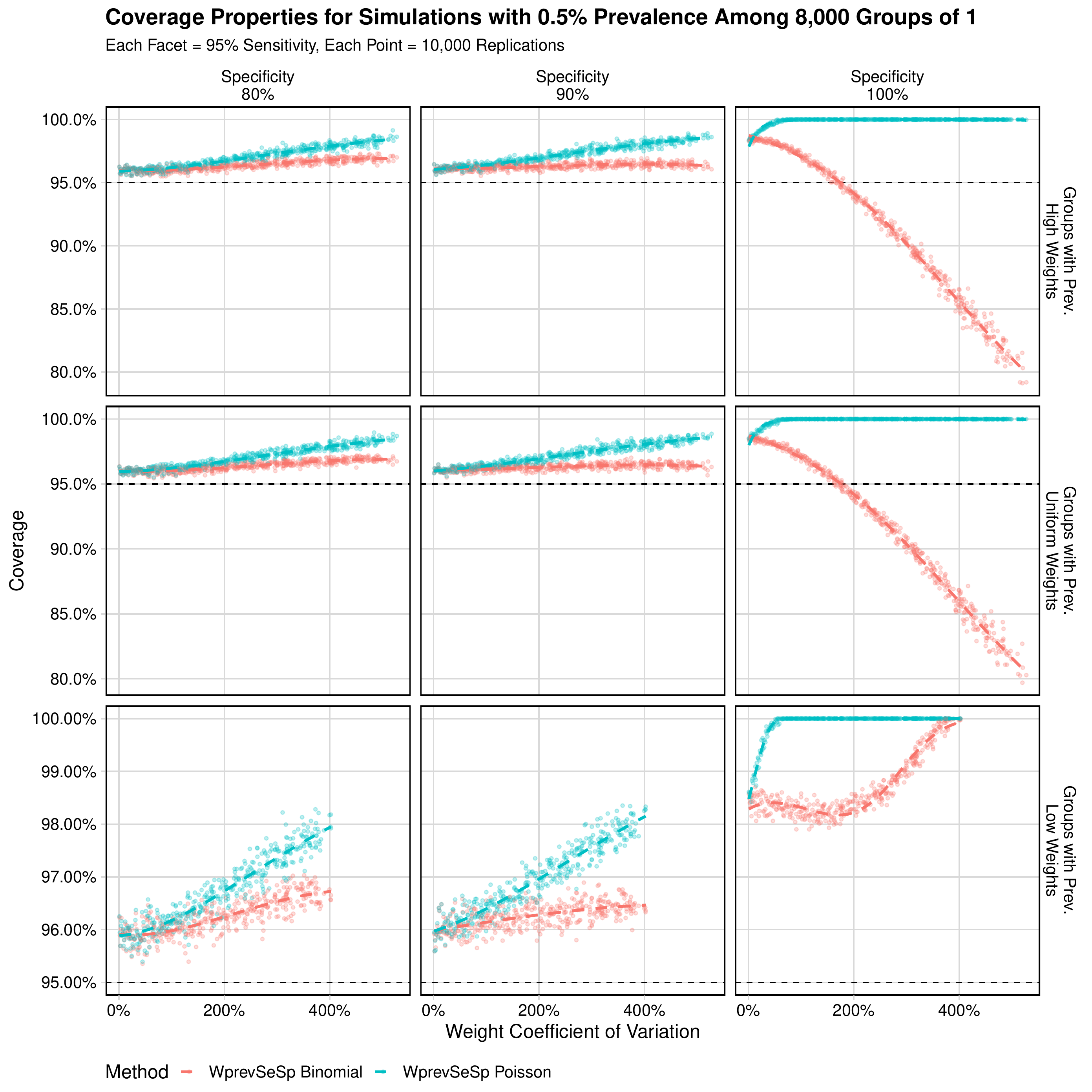}
\caption{Coverage properties for the confidence interval procedures, WprevSeSp Binomial and WprevSeSp Poisson.
Each point represents 10,000 simulations of datasets from a population with 0.5\% Prevalence where 8000 individuals are sampled.
Each datasets also includes simulated results of tests to evaluate the sensitivity and specificity of the assay performed on 60 and 300 individuals, respectively.
The horizontal dashed line indicates the nominal coverage, 95\%.
Colored dashed lines are estimates from a logistic regression model using quadratic splines.}
\label{fig:imperfect_coverage_8000_groups_0_005_prev}
\end{figure}

\begin{figure}
\centering
\includegraphics[width=0.8\textwidth]{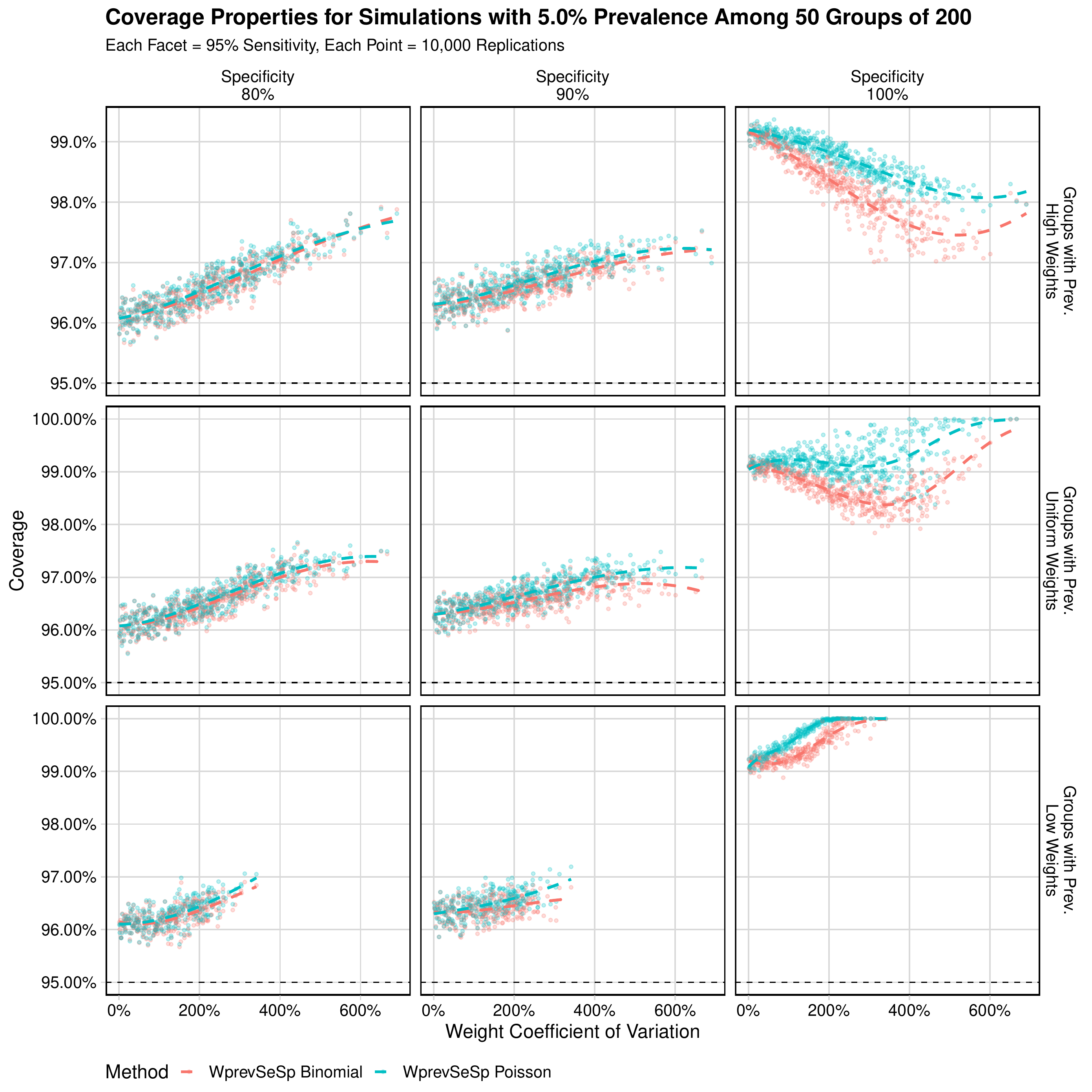}
\caption{Coverage properties for the confidence interval procedures, WprevSeSp Binomial and WprevSeSp Poisson.
Each point represents 10,000 simulations of datasets from a population with 5\% Prevalence where 50 groups of 200 people are sampled.
Each datasets also includes simulated results of tests to evaluate the sensitivity and specificity of the assay performed on 60 and 300 individuals, respectively.
The horizontal dashed line indicates the nominal coverage, 95\%.
Colored dashed lines are estimates from a logistic regression model using quadratic splines.}
\label{fig:imperfect_coverage_50_groups_0_05_prev}
\end{figure}

\begin{figure}
\centering
\includegraphics[width=0.8\textwidth]{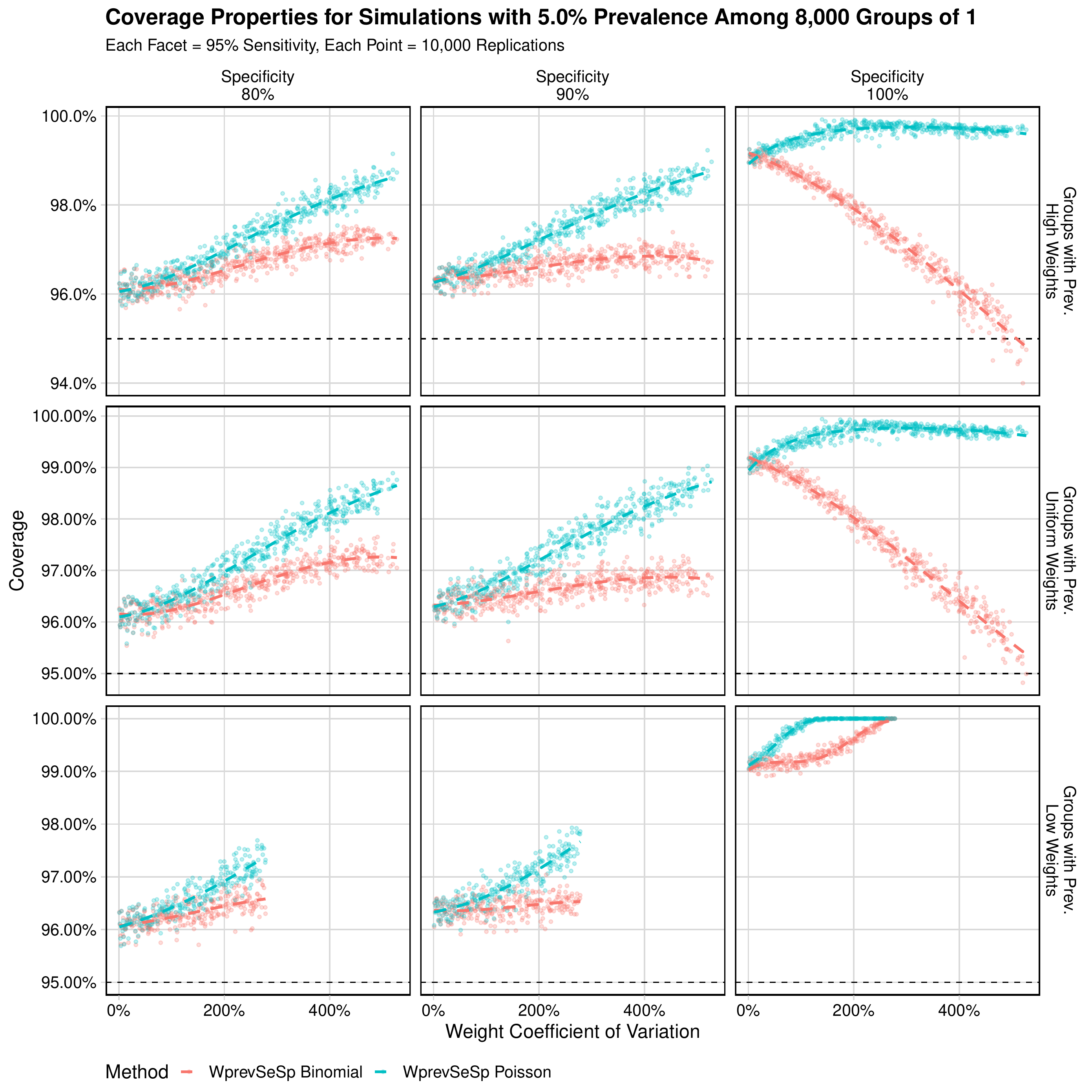}
\caption{Coverage properties for the confidence interval procedures, WprevSeSp Binomial and WprevSeSp Poisson.
Each point represents 10,000 simulations of datasets from a population with 5\% Prevalence where 8000 individuals are sampled.
Each datasets also includes simulated results of tests to evaluate the sensitivity and specificity of the assay performed on 60 and 300 individuals, respectively.
The horizontal dashed line indicates the nominal coverage, 95\%.
Colored dashed lines are estimates from a logistic regression model using quadratic splines.}
\label{fig:imperfect_coverage_8000_groups_0_05_prev}
\end{figure}

The coverage properties for these simulations are presented in Figures~\ref{fig:imperfect_coverage_50_groups_0_005_prev}--\ref{fig:imperfect_coverage_8000_groups_0_05_prev}.
Additional properties for these simulations are presented in Figures~\ref{fig:imperfect_lower_error_frequency_50_groups_0_05_prev}--\ref{fig:imperfect_confidence_interval_width_8000_groups_0_05_prev}.

Based on Figures~\ref{fig:imperfect_lower_error_frequency_50_groups_0_05_prev}--\ref{fig:imperfect_upper_error_frequency_8000_groups_0_05_prev}, we note that, in most settings, the two melding procedures result in conservative confidence intervals, often nearing 100\% coverage.
With perfect specificity, the WprevSeSp Binomial method fails to maintain nominal coverage when the coefficient of variation among the weights is high and specificity is perfect.
Only our proposed WprevSeSp Poisson method maintains or exceed the desired coverage in all scenarios.
In these scenarios, we also assess properties of the wsPoisson procedure, which does not account for the the imperfect assay.
This method results in approximately 0\% coverage in any scenarios where specificity is less than perfect.
For this reason, results from method are omitted in the figures.
Because our methods appear to be very conservative, with coverage near 100\% in some cases, we present the widths of the confidence intervals in Figures~\ref{fig:imperfect_confidence_interval_width_50_groups_0_005_prev}--\ref{fig:imperfect_confidence_interval_width_8000_groups_0_05_prev}.
The WprevSeSp Binomial and  WprevSeSp Poisson methods typically produce wide intervals of approximately the same size - sometimes as wide as 12\%, even when true prevalence is 0.05\%.
One notable exception to this is presented in Figure \ref{fig:imperfect_confidence_interval_width_8000_groups_0_005_prev}, which shows that for tests with perfect specificity, the WprevSeSP Binomial method produces much narrower confidence intervals than the other method.

\section{Applications}
\label{sec-Application}

We  apply these two methods to a real data set from
Kalish, et al \cite{Kali:2021}.
This data set was collected to estimate seroprevalence of SARS-CoV-2 in undiagnosed adults in the United States between May and July 2020.
The assay used in this data is estimated to have perfect sensitivity, based on 56 tests on individuals with confirmed SARS-CoV-2 and perfect specificity based on 300 tests on individuals confirmed to not have SARS-CoV-2.
First we apply the methods to the full data set (\( n =  8058, \text{weight coefficient of variation} = 252\%\)).
The seroprevalence in Kalish, et al was 4.6\% with (95\% CI: 2.6\% to 6.5\%), using a confidence interval method that was nearly the same as the WprevSeSp Binomial method (the method Kalish et al  included a calculation of the variability of the weights due to the estimation of the weights, whereas in this paper we treat the weights as fixed constants).   
The Korn and Graubard type melded confidence interval with imperfect assay adjustments (WprevSeSp Binomial) studied in this paper produced the 95\% confidence interval for population prevalence nearly the equivalent, (2.53\%, 6.68\%). while the wsPoisson type melded confidence interval with imperfect assay adjustments (WprevSeSp Poisson) produced the 95\% confidence interval (2.56\%, 7.54\%).
We also apply the wsPoisson method from Section~\ref{sec:weight-perfect}, which does not account for imperfections in the assay, resulting in a 95\% confidence interval of (3.04\%, 7.39\%).
While all three intervals overlap to a large degree, the WprevSeSp Poisson interval is the widest.
Our simulations show that in this situation, the WprevSeSp Binomial interval may be the best, because with coefficient of variance about 250\% (see Figures~B15 and B20, top right panel) the error on both sides of the confidence interval is bounded at 2.5\% and the width of the intervals are better (Figure~B23).

We also apply the methods to the subset of only Hispanic participants (\( n = 1281, \text{weight coefficient of variation} = 306\% \)), where Kalish et al estimated the undiagnosed adult seroprevalence estimate as 6.1\% (95\% CI: 2.4\% to 11.5\%).
The WprevSeSp Binomial method produces a 95\% confidence interval for population prevalence (2.35\%, 11.75\%), while the WprevSeSp Poisson method produces a 95\% confidence interval (2.40\%, 20.02\%).
The wsPoisson method produces a 95\% confidence interval of (2.80\%, 19.63\%).
In this case, the two melded confidence intervals are much wider than the WprevSeSp Binomial interval, which is as expected since the melding method is designed to guarantee coverage, although the simulations show that the WprevSeSp Binomial interval may be reasonable (see e.g., Figure~B20).
The smaller WprevSeSp Binomial interval is also unsurprising and is similar to the results observed in our simulation study.

\pagebreak

\section{Discussion}

We presented several methods for creating confidence intervals to assess disease prevalence in variety of settings, including simple random samples with imperfect tests, weighted sampling with perfect tests, and weighted sampling with imperfect tests.
One of the new methods was very similar to the method used by Kalish et al\cite{Kali:2021}, and in this paper we have explored its properties.
These new confidence intervals appear to guarantee coverage in most simulated settings, and in general to demonstrate higher coverage than competitor methods.
In the case of the simple random sample with an imperfect test, our new methods are able to bound the lower error rate for a 95\% confidence interval at 2.5\%, while the Lang-Reiczigel\cite{Lang:2014} method maintains 95\% coverage by allowing a higher lower error rate.
A big advantage of our method is that it may be applied with complex survey methods, where each individual has their own weight, such as in Kalish et al\cite{Kali:2021}. 
However, our method  only studied fixed weights and not when the weights are estimated as in Kalish et al.  Further worked is needed to address such cases.  In addition, further worked is needed to consider other complex sample designs such as multistage cluster designs that are used in  household and institutional surveys such hospital and medical practice surveys.

Our method was slightly different to that used in Kalish, et al\cite{Kali:2021}, in that thdatae latter method included estimates of the variability of the weights; however, recalculating the confidence intervals on the same data shows that in that case there was little difference. 

Our methods' conservative properties are especially advantageous in settings where the competitor methods exhibit much lower than nominal coverage.
For example, there is high variance among the sample weights and prevalence is concentrated among the highest-weighted samples competitor coverage of 95\% confidence intervals can fall to 60\% for competitor methods, while our method exhibits > 95\% coverage.
Thus, we suggest that our melding methods be employed when working with survey settings which involve high variance among the weights or lower errors are particularly undesirable.

\section{Acknowledgements}
For sharing the data from the Kalish, et al study, we thank Matthew Memoli and the LID Clinical Studies Unit of the National Institute of Allergy and Infectious Diseases, NIH,  Kaitlyn Sadtler from the National Institute of Biomedical Imaging and Bioengineering, NIH,   Matthew Hall from the National Center for Advancing Translational Sciences, NIH, and Dominic Esposito from the Fredrick National Laboratory for Cancer Research, NCI, NIH.

\section{Data Availability Statement}
Data sharing is not applicable to this article as no new data were created or analyzed in this study.

\section{Bibliography}
\nocite{*}
\bibliography{refs}%

\begin{thebibliography}{10}
\providecommand \doibase [0]{http://dx.doi.org/}%

\bibitem{hemenwaySelfDefense}
Hemenway D. The Myth of Millions of Annual Self-Defense Gun Uses: A Case Study
  of Survey Overestimates of Rare Events. {\it CHANCE} 1997\string;
  10(3)\string: 6-10.
\newblock \href {\doibase 10.1080/09332480.1997.10542033} {doi:
  10.1080/09332480.1997.10542033}

\bibitem{Dean:2015}
Dean N, Pagano M. Evaluating confidence interval methods for binomial
  proportions in clustered surveys. {\it Journal of Survey Statistics and
  Methodology} 2015\string; 3(4)\string: 484--503.

\bibitem{franco2019}
Franco C, Little RJ, Louis TA, Slud EV. Comparative study of confidence
  intervals for proportions in complex sample surveys. {\it Journal of survey
  statistics and methodology} 2019\string; 7(3)\string: 334--364.

\bibitem{Lang:2014}
Lang Z, Reiczigel J. Confidence limits for prevalence of disease adjusted for
  estimated sensitivity and specificity. {\it Preventive Veterinary Medicine}
  2014\string; 113(1)\string: 13-22.
\newblock \href {\doibase https://doi.org/10.1016/j.prevetmed.2013.09.015}
  {doi: https://doi.org/10.1016/j.prevetmed.2013.09.015}

\bibitem{DiCi:2021}
DiCiccio TJ, Ritzwoller DM, Romano JP, Shaikh AM. Confidence Intervals for
  Seroprevalence.  2021.
\newblock \href {\doibase 10.48550/ARXIV.2103.15018} {doi:
  10.48550/ARXIV.2103.15018}

\bibitem{Cai:2020}
Cai B, Ioannidis JPA, Bendavid E, Tian L. Exact inference for disease
  prevalence based on a test with unknown specificity and sensitivity. {\it
  Journal of Applied Statistics} 2022\string; 0(0)\string: 1-25.
\newblock \href {\doibase 10.1080/02664763.2021.2019687} {doi:
  10.1080/02664763.2021.2019687}

\bibitem{Berg:1994}
Berger RL, Boos DD. P Values Maximized Over a Confidence Set for the Nuisance
  Parameter. {\it Journal of the American Statistical Association} 1994\string;
  89(427)\string: 1012-1016.
\newblock \href {\doibase 10.1080/01621459.1994.10476836} {doi:
  10.1080/01621459.1994.10476836}

\bibitem{Kali:2021}
Kalish H, Klumpp-Thomas C, Hunsberger S, et al. Undiagnosed SARS-CoV-2
  Seropositivity During the First Six Months of the COVID-19 Pandemic in the
  United States. {\it Science Translational Medicine} 2021.

\bibitem{rosin2021estimating}
Rosin S, Shook-Sa BE, Cole SR, Hudgens MG. Estimating SARS-CoV-2
  Seroprevalence.;  2021.

\bibitem{GelmanBayes}
Gelman A, Carpenter B. Bayesian analysis of tests with unknown specificity and
  sensitivity. {\it Journal of the Royal Statistical Society: Series C (Applied
  Statistics)} 2020\string; 69(5)\string: 1269-1283.
\newblock \href {\doibase https://doi.org/10.1111/rssc.12435} {doi:
  https://doi.org/10.1111/rssc.12435}

\bibitem{FayP:2015}
Fay MP, Proschan MA, Brittain E. Combining one-sample confidence procedures for
  inference in the two-sample case. {\it Biometrics} 2015\string; 71(1)\string:
  146--156.

\bibitem{Xie2013}
Xie Mg, Singh K. Confidence Distribution, the Frequentist Distribution
  Estimator of a Parameter: A Review. {\it International Statistical Review}
  2013\string; 81(1)\string: 3-39.
\newblock \href {\doibase https://doi.org/10.1111/insr.12000} {doi:
  https://doi.org/10.1111/insr.12000}

\bibitem{Fay2021}
Fay MP, Proschan MA, Brittain E, Tiwari R. Interpreting P-values and Confidence
  Intervals using Well-Calibrated Null Preference Priors. {\it Statistical
  Science}\string: to appear.

\bibitem{10.1093/biomet/26.4.404}
Clopper CJ, Pearson ES. {The Use of Confidence or Fiducial Limits Illustrated
  in the Case of the Binomial}. {\it Biometrika} 1934\string; 26(4)\string:
  404-413.
\newblock \href {\doibase 10.1093/biomet/26.4.404} {doi:
  10.1093/biomet/26.4.404}

\bibitem{asht}
Fay MP. {\it asht: Applied Statistical Hypothesis Tests}. ; :  2020.
\newblock R package version 0.9.6.

\bibitem{FayF:1997}
Fay MP, Feuer EJ. Confidence intervals for directly standardized rates: a
  method based on the gamma distribution. {\it Statistics in medicine}
  1997\string; 16(7)\string: 791--801.

\bibitem{AgrestiCoull}
Agresti A, Coull BA. Approximate is Better than “Exact” for Interval
  Estimation of Binomial Proportions. {\it The American Statistician}
  1998\string; 52(2)\string: 119-126.
\newblock \href {\doibase 10.1080/00031305.1998.10480550} {doi:
  10.1080/00031305.1998.10480550}

\bibitem{Korn:1998}
Korn EL, Graubard BI. Confidence intervals for proportions with small expected
  number of positive counts estimated from survey data. {\it Survey
  Methodology} 1998\string; 24\string: 193--201.

\bibitem{Korn:1999}
Korn EL, Graubard BI. {\it Analysis of health surveys}.
\newblock John Wiley \& Sons .
\newblock 1999.

\bibitem{Baker:1994}
Baker SG. The multinomial-Poisson transformation. {\it Journal of the Royal
  Statistical Society: Series D (The Statistician)} 1994\string; 43(4)\string:
  495--504.

\bibitem{FayK:2017}
Fay MP, Kim S. Confidence intervals for directly standardized rates using mid-p
  gamma intervals. {\it Biometrical Journal} 2017\string; 59(2)\string:
  377--387.

\bibitem{Roga:1978}
Rogan WJ, Gladen B. Estimating prevalence from the results of a screening test.
  {\it American journal of epidemiology} 1978\string; 107(1)\string: 71--76.

\bibitem{Lemeshow1985SurveysTM}
Lemeshow S, Robinson D. Surveys to measure programme coverage and impact: a
  review of the methodology used by the expanded programme on immunization..
  {\it World health statistics quarterly. Rapport trimestriel de statistiques
  sanitaires mondiales} 1985\string; 38(1)\string: 65-75.

\bibitem{ClopperPearson}
Clopper CJ, Pearson ES. {The Use of Confidence or Fiducial Limits Illustrated
  in the Case of the Binomial}. {\it Biometrika} 1934\string; 26(4)\string:
  404-413.
\newblock \href {\doibase 10.1093/biomet/26.4.404} {doi:
  10.1093/biomet/26.4.404}

\bibitem{gustafson2003measurement}
Gustafson P. {\it Measurement error and misclassification in statistics and
  epidemiology: impacts and Bayesian adjustments}.
\newblock CRC Press .
\newblock 2003.

\bibitem{Flor2020}
Flor M, Wei{\ss} M, Selhorst T, M{\"u}ller-Graf C, Greiner M. Comparison of
  Bayesian and frequentist methods for prevalence estimation under
  misclassification. {\it BMC Public Health} 2020\string; 20(1)\string: 1135.
\newblock \href {\doibase 10.1186/s12889-020-09177-4} {doi:
  10.1186/s12889-020-09177-4}

\end{thebibliography}

\appendix

\section{Monotonicity of \( g \)}
\label{monotonicity}
It is clear that \( g \) is monotonic within each of its piecewise-defined functions.
In the following sections we consider if monotonicity holds at the change points between piecewise functions.

\subsection{Monotonicity in \( \hat{\phi}_p \)}

\begin{itemize}
	\item Case 1: \( \hat{\phi}_n < \hat{\theta}_1 \). Not monotonic. \\
\begin{equation} 
g(\hat{\theta}_1, \hat{\phi}_n, \hat{\phi}_p)
=
\left\{ 
\begin{array}{ll}
0 & \hat{\phi}_p < \hat{\phi}_n < \hat{\theta}_1  \\
0 & \hat{\phi}_n = \hat{\phi}_p < \hat{\theta}_1 \\
1 &	\hat{\phi}_n < \hat{\phi}_p < \hat{\theta}_1 \\
\frac{\hat{\theta}_1 - \hat{\phi}_n}{\hat{\phi}_p - \hat{\phi}_n} = 1 & \hat{\phi}_n < \hat{\phi}_p = \hat{\theta}_1 \\
\frac{\hat{\theta}_1 - \hat{\phi}_n}{\hat{\phi}_p - \hat{\phi}_n} & \hat{\phi}_n < \hat{\theta}_1 < \hat{\phi}_p \\
\end{array}
\right.
\end{equation}

\item Case 2: \( \hat{\theta}_1 < \hat{\phi}_n \). Monotonic. \\
\begin{equation} 
g(\hat{\theta}_1, \hat{\phi}_n, \hat{\phi}_p)
=
\left\{ 
\begin{array}{ll}
0 & \hat{\phi}_p < \hat{\theta}_1 < \hat{\phi}_n  \\
0 & \hat{\theta}_1 = \hat{\phi}_p < \hat{\phi}_n \\
0 &	\hat{\theta}_1 < \hat{\phi}_p < \hat{\phi}_n \\
0 & \hat{\theta}_1 < \hat{\phi}_p = \hat{\phi}_n \\
0 & \hat{\theta}_1 < \hat{\phi}_n < \hat{\phi}_p \\
\end{array}
\right.
\end{equation}

\item Case 3: \( \hat{\theta}_1 = \hat{\phi}_n \). Monotonic. \\
\begin{equation} 
g(\hat{\theta}_1, \hat{\phi}_n, \hat{\phi}_p)
=
\left\{ 
\begin{array}{ll}
0 & \hat{\phi}_p < \hat{\theta}_1 = \hat{\phi}_n	\\
\frac{\hat{\theta}_1 - \hat{\phi}_n}{\hat{\phi}_p - \hat{\phi}_n} = \frac{0}{0} = 0 & \hat{\phi}_p = \hat{\theta}_1 = \hat{\phi}_n 	\\
\frac{\hat{\theta}_1 - \hat{\phi}_n}{\hat{\phi}_p - \hat{\phi}_n} = \frac{0}{\hat{\phi}_p - \hat{\phi}_n} = 0 & \hat{\theta}_1 = \hat{\phi}_n < \hat{\phi}_p	\\
\end{array}
\right.
\end{equation}
\end{itemize}

\subsection{Monotonicity in \( \hat{\theta}_1 \)}

\begin{itemize}
	\item Case 1: \( \hat{\phi}_n < \hat{\phi}_p \). Monotonic. \\
\begin{equation} 
g(\hat{\theta}_1, \hat{\phi}_n, \hat{\phi}_p)
=
\left\{ 
\begin{array}{ll}
0 & \hat{\theta}_1 < \hat{\phi}_n < \hat{\phi}_p  \\
\frac{\hat{\theta}_1 - \hat{\phi}_n}{\hat{\phi}_p - \hat{\phi}_n} = 0 & \hat{\phi}_n = \hat{\theta}_1 < \hat{\phi}_p \\
\frac{\hat{\theta}_1 - \hat{\phi}_n}{\hat{\phi}_p - \hat{\phi}_n} &	\hat{\phi}_n < \hat{\theta}_1 < \hat{\phi}_p \\
\frac{\hat{\theta}_1 - \hat{\phi}_n}{\hat{\phi}_p - \hat{\phi}_n} = 1 & \hat{\phi}_n < \hat{\theta}_1 = \hat{\phi}_p \\
1 & \hat{\phi}_n < \hat{\phi}_p < \hat{\theta}_1 \\
\end{array}
\right.
\end{equation}

\item Case 2: \( \hat{\phi}_p < \hat{\phi}_n \). Monotonic. \\
\begin{equation} 
g(\hat{\theta}_1, \hat{\phi}_n, \hat{\phi}_p)
=
\left\{ 
\begin{array}{ll}
0 & \hat{\theta}_1 < \hat{\phi}_p < \hat{\phi}_n  \\
0 & \hat{\phi}_p = \hat{\theta}_1 < \hat{\phi}_n \\
0 &	\hat{\phi}_p < \hat{\theta}_1 < \hat{\phi}_n \\
0 & \hat{\phi}_p < \hat{\theta}_1 = \hat{\phi}_n \\
0 & \hat{\phi}_p < \hat{\phi}_n < \hat{\theta}_1 \\
\end{array}
\right.
\end{equation}

\item Case 3: \( \hat{\phi}_p = \hat{\phi}_n \). Monotonic. \\
\begin{equation} 
g(\hat{\theta}_1, \hat{\phi}_n, \hat{\phi}_p)
=
\left\{ 
\begin{array}{ll}
0 & \hat{\theta}_1 < \hat{\phi}_p = \hat{\phi}_n	\\
\frac{\hat{\theta}_1 - \hat{\phi}_n}{\hat{\phi}_p - \hat{\phi}_n} = \frac{0}{0} = 0 & \hat{\theta}_1 = \hat{\phi}_p = \hat{\phi}_n 	\\
0 & \hat{\phi}_p = \hat{\phi}_n < \hat{\theta}_1	\\
\end{array}
\right.
\end{equation}
\end{itemize}

\subsection{Monotonicity in \( \hat{\phi}_n \)}

\begin{itemize}
	\item Case 1: \( \hat{\phi}_p < \hat{\theta}_1 \). Monotonic. \\
\begin{equation} 
g(\hat{\theta}_1, \hat{\phi}_n, \hat{\phi}_p)
=
\left\{ 
\begin{array}{ll}
1 & \hat{\phi}_n < \hat{\phi}_p < \hat{\theta}_1  \\
0 & \hat{\phi}_p = \hat{\phi}_n < \hat{\theta}_1 \\
0 &	\hat{\phi}_p < \hat{\phi}_n < \hat{\theta}_1 \\
0 & \hat{\phi}_p < \hat{\phi}_n = \hat{\theta}_1 \\
0 & \hat{\phi}_p < \hat{\theta}_1 < \hat{\phi}_n \\
\end{array}
\right.
\end{equation}

\item Case 2: \( \hat{\theta}_1 < \hat{\phi}_p \). Monotonic.
\begin{equation} 
g(\hat{\theta}_1, \hat{\phi}_n, \hat{\phi}_p)
=
\left\{ 
\begin{array}{ll}
\frac{\hat{\theta}_1 - \hat{\phi}_n}{\hat{\phi}_p - \hat{\phi}_n} & \hat{\phi}_n < \hat{\theta}_1 < \hat{\phi}_p  \\
\frac{\hat{\theta}_1 - \hat{\phi}_n}{\hat{\phi}_p - \hat{\phi}_n} = \frac{0}{\hat{\phi}_p - \hat{\phi}_n} = 0 & \hat{\theta}_1 = \hat{\phi}_n < \hat{\phi}_p \\
0 &	\hat{\theta}_1 < \hat{\phi}_n < \hat{\phi}_p \\
0 & \hat{\theta}_1 < \hat{\phi}_n = \hat{\phi}_p \\
0 & \hat{\theta}_1 < \hat{\phi}_p < \hat{\phi}_n \\
\end{array}
\right.
\end{equation}

\item Case 3: \( \hat{\theta}_1 = \hat{\phi}_p \). Monotonic. \\
\begin{equation} 
g(\hat{\theta}_1, \hat{\phi}_n, \hat{\phi}_p)
=
\left\{ 
\begin{array}{ll}
\frac{\hat{\theta}_1 - \hat{\phi}_n}{\hat{\phi}_p - \hat{\phi}_n} = 1 & \hat{\phi}_n < \hat{\theta}_1 = \hat{\phi}_p	\\
\frac{\hat{\theta}_1 - \hat{\phi}_n}{\hat{\phi}_p - \hat{\phi}_n} = \frac{0}{0} = 0 & \hat{\phi}_n = \hat{\theta}_1 = \hat{\phi}_p 	\\
0 & \hat{\theta}_1 = \hat{\phi}_p < \hat{\phi}_n	\\
\end{array}
\right.
\end{equation}
\end{itemize}

\section{Additional Figures}

\begin{figure}
\centering
\includegraphics[width=0.8\textwidth]{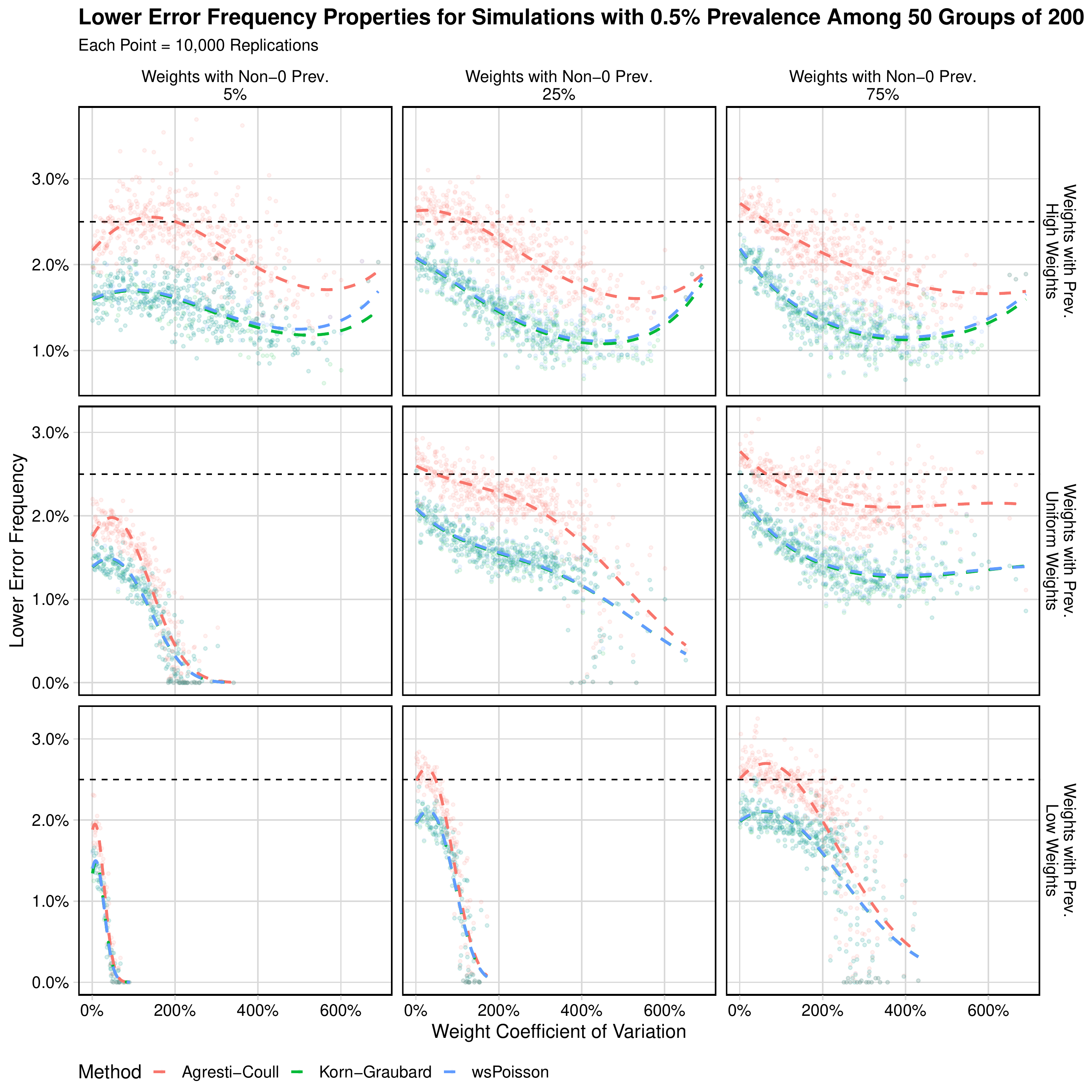}
\caption{Lower error properties for the wsPoisson model and two standard methods, the Dean-Pagano modification of the Agresti-Coull method and of the Korn-Graubard method.
Each point represents 10,000 simulations of datasets from a population with 0.5\% Prevalence where 50 groups of 200 people are sampled.
The horizontal dashed line indicates the nominal lower error rate, 2.5\%.
Colored dashed lines are estimates from a logistic regression model using quadratic splines.}
\label{fig:perfect_lower_error_frequency_50_groups_0_005_prev}
\end{figure}

\begin{figure}
\centering
\includegraphics[width=0.8\textwidth]{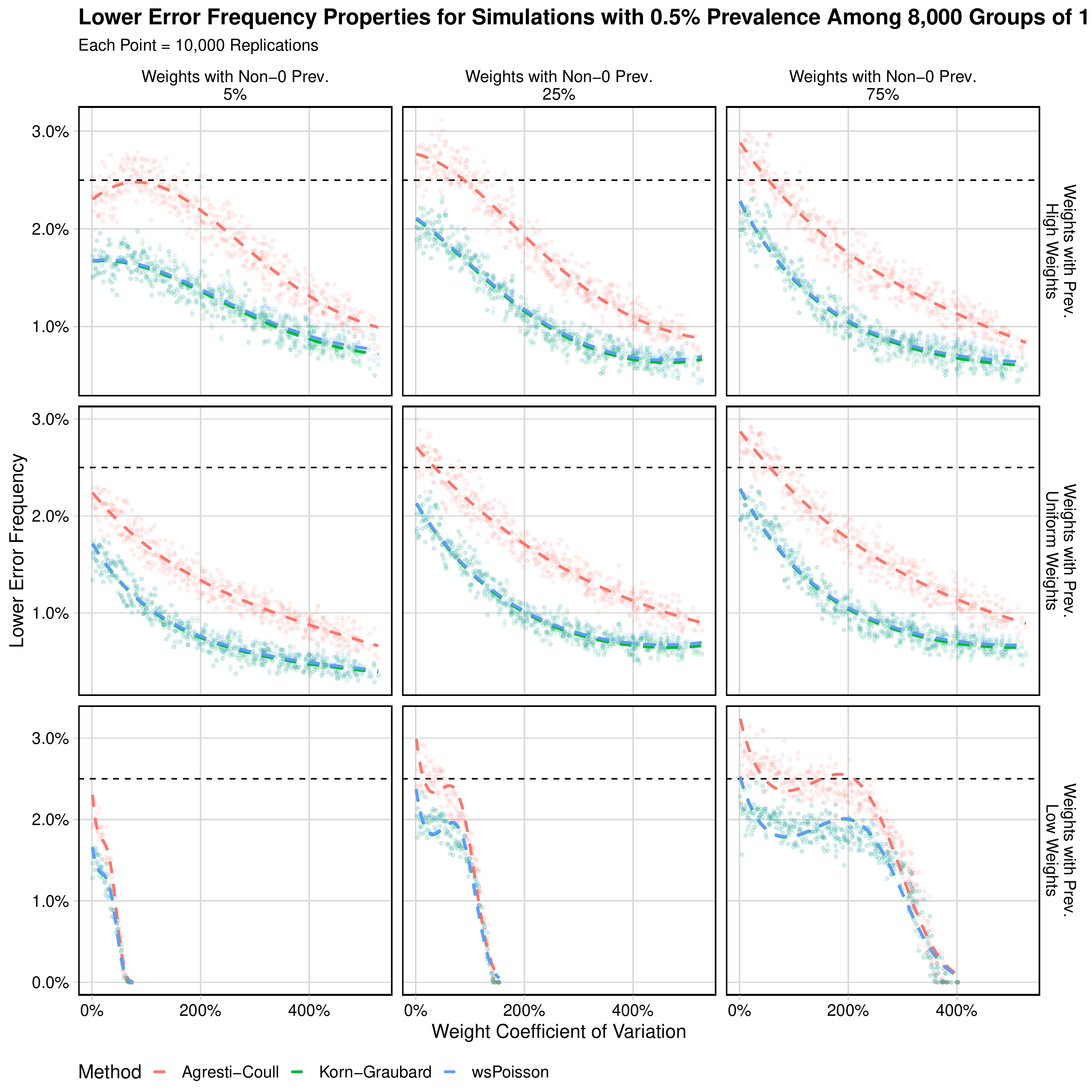}
\caption{Lower error properties for the wsPoisson model and two standard methods, the Dean-Pagano modification of the Agresti-Coull method and of the Korn-Graubard method.
Each point represents 10,000 simulations of datasets from a population with 0.5\% Prevalence where 8000 individuals are sampled.
The horizontal dashed line indicates the nominal lower error rate, 2.5\%.
Colored dashed lines are estimates from a logistic regression model using quadratic splines.}
\label{fig:perfect_lower_error_frequency_8000_groups_0_005_prev}
\end{figure}

\begin{figure}
\centering
\includegraphics[width=0.8\textwidth]{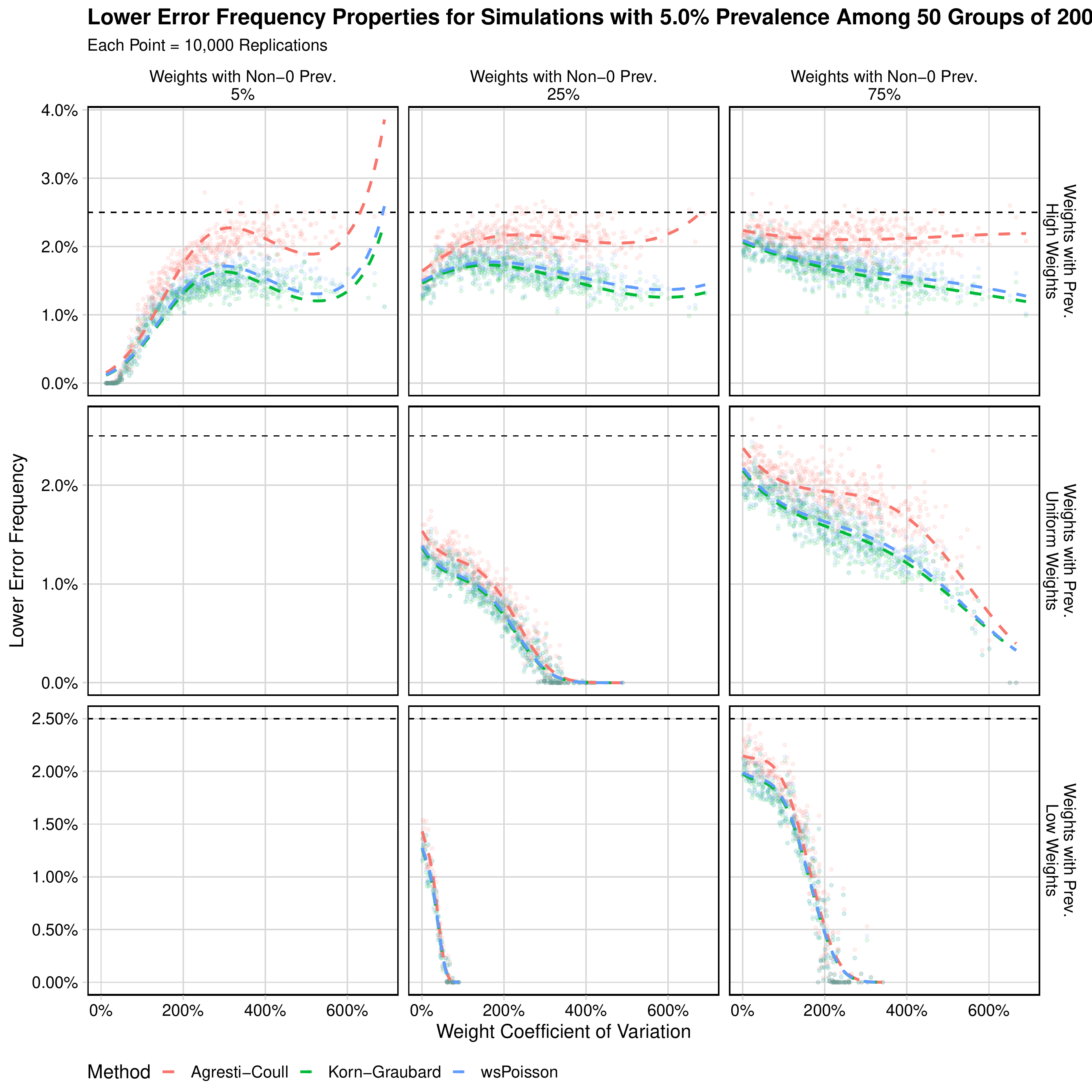}
\caption{Lower error properties for the wsPoisson model and two standard methods, the Dean-Pagano modification of the Agresti-Coull method and of the Korn-Graubard method.
Each point represents 10,000 simulations of datasets from a population with 5\% Prevalence where 50 groups of 200 people are sampled.
The horizontal dashed line indicates the nominal lower error rate, 2.5\%.
Colored dashed lines are estimates from a logistic regression model using quadratic splines.}
\label{fig:perfect_lower_error_frequency_50_groups_0_05_prev}
\end{figure}

\begin{figure}
\centering
\includegraphics[width=0.8\textwidth]{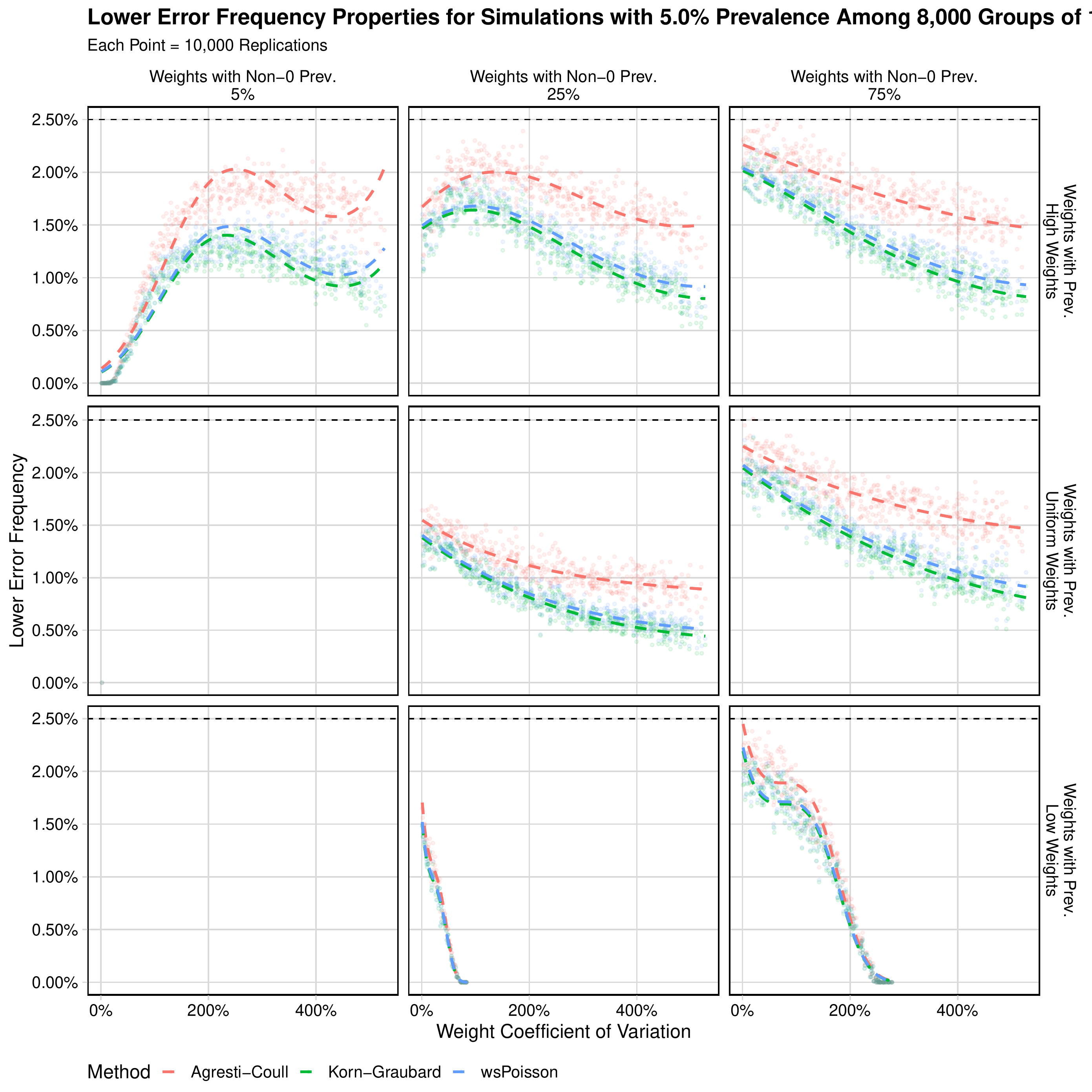}
\caption{Lower error properties for the wsPoisson model and two standard methods, the Dean-Pagano modification of the Agresti-Coull method and of the Korn-Graubard method.
Each point represents 10,000 simulations of datasets from a population with 5\% Prevalence where 8000 individuals are sampled.
The horizontal dashed line indicates the nominal lower error rate, 2.5\%.
Colored dashed lines are estimates from a logistic regression model using quadratic splines.}
\label{fig:perfect_lower_error_frequency_8000_groups_0_05_prev}
\end{figure}

\begin{figure}
\centering
\includegraphics[width=0.8\textwidth]{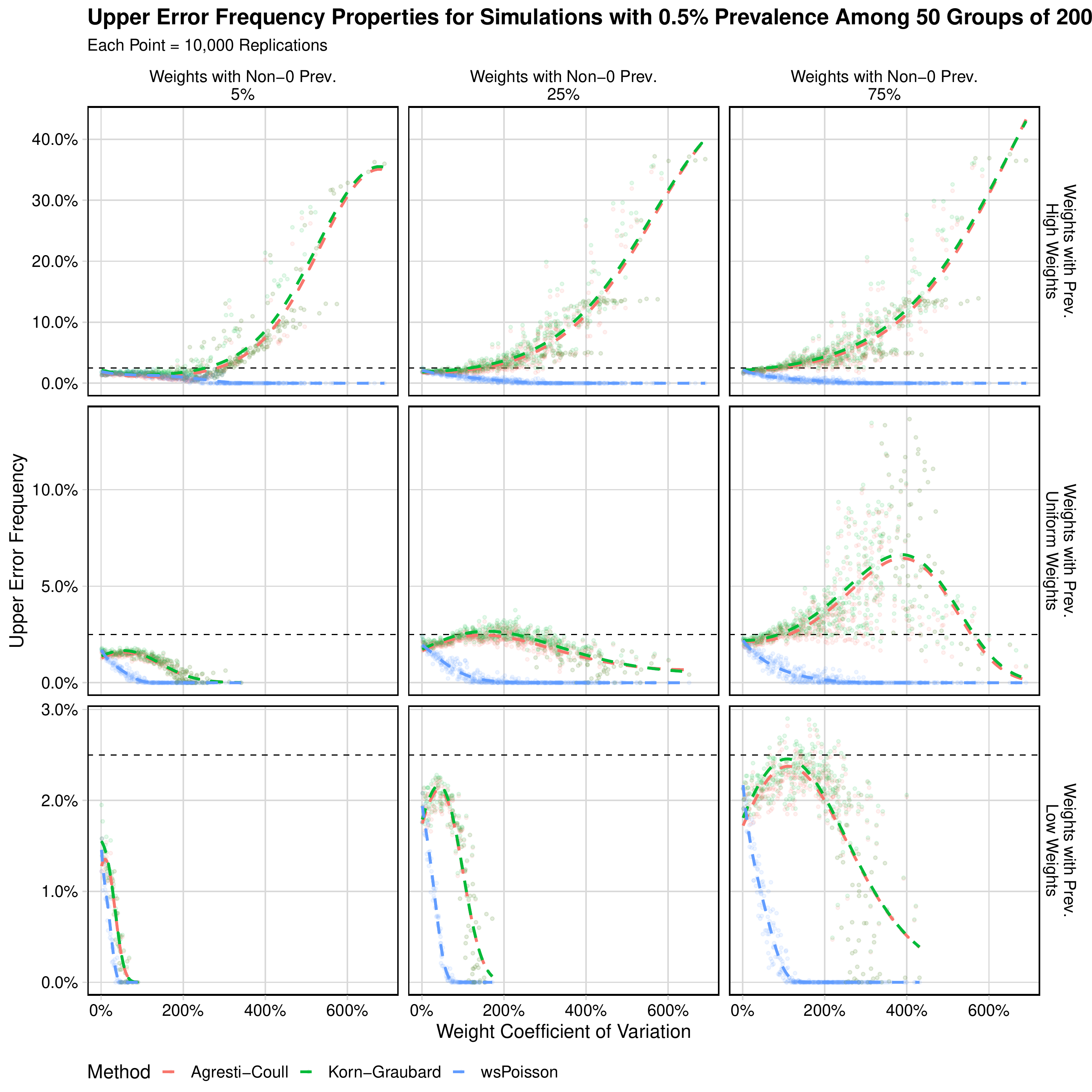}
\caption{Upper error properties for the wsPoisson model and two standard methods, the Dean-Pagano modification of the Agresti-Coull method and of the Korn-Graubard method.
Each point represents 10,000 simulations of datasets from a population with 0.5\% Prevalence where 50 groups of 200 people are sampled.
The horizontal dashed line indicates the nominal upper error rate, 2.5\%.
Colored dashed lines are estimates from a logistic regression model using quadratic splines.}
\label{fig:perfect_upper_error_frequency_50_groups_0_005_prev}
\end{figure}

\begin{figure}
\centering
\includegraphics[width=0.8\textwidth]{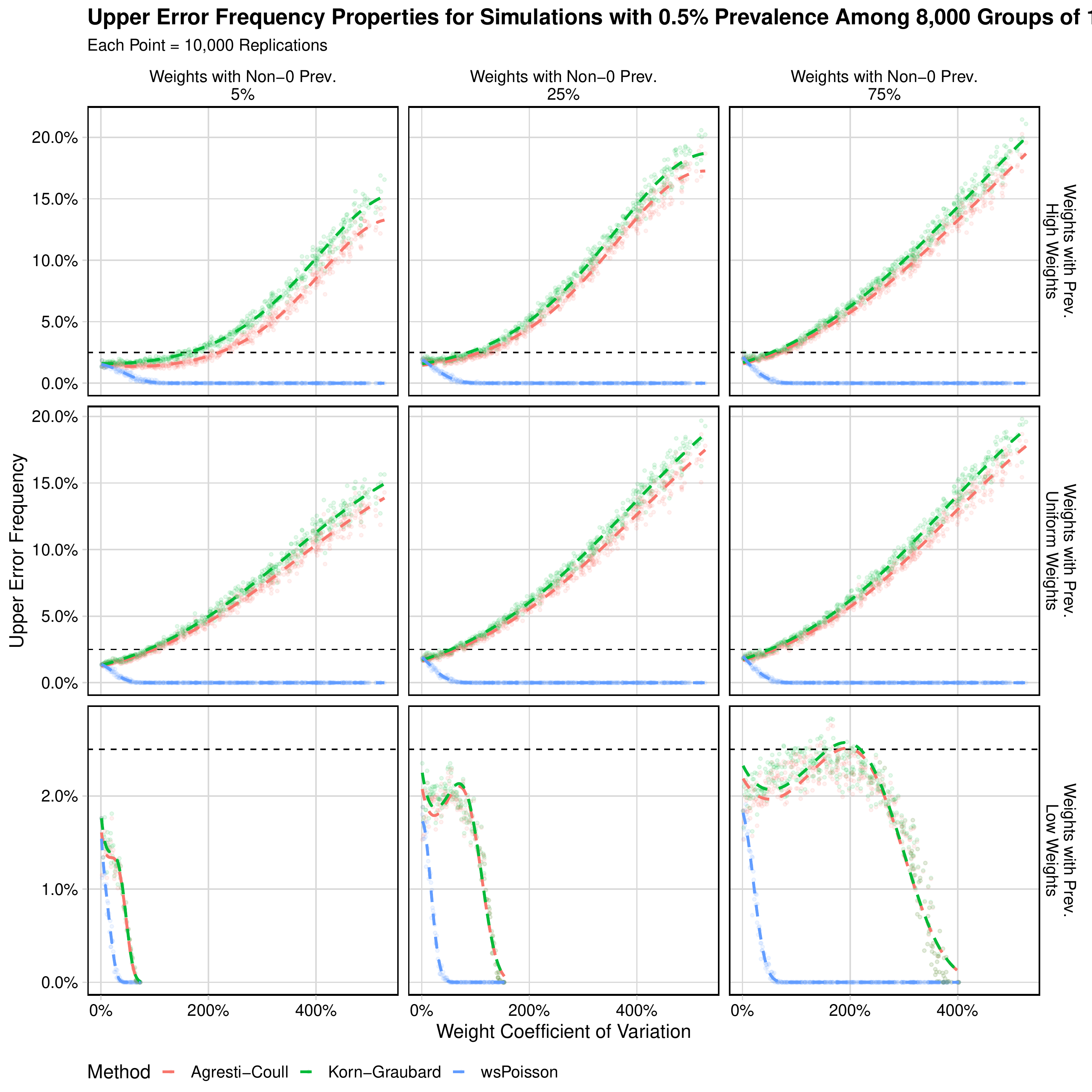}
\caption{Upper error properties for the wsPoisson model and two standard methods, the Dean-Pagano modification of the Agresti-Coull method and of the Korn-Graubard method.
Each point represents 10,000 simulations of datasets from a population with 0.5\% Prevalence where 8000 individuals are sampled.
The horizontal dashed line indicates the nominal upper error rate, 2.5\%.
Colored dashed lines are estimates from a logistic regression model using quadratic splines.}
\label{fig:perfect_upper_error_frequency_8000_groups_0_005_prev}
\end{figure}

\begin{figure}
\centering
\includegraphics[width=0.8\textwidth]{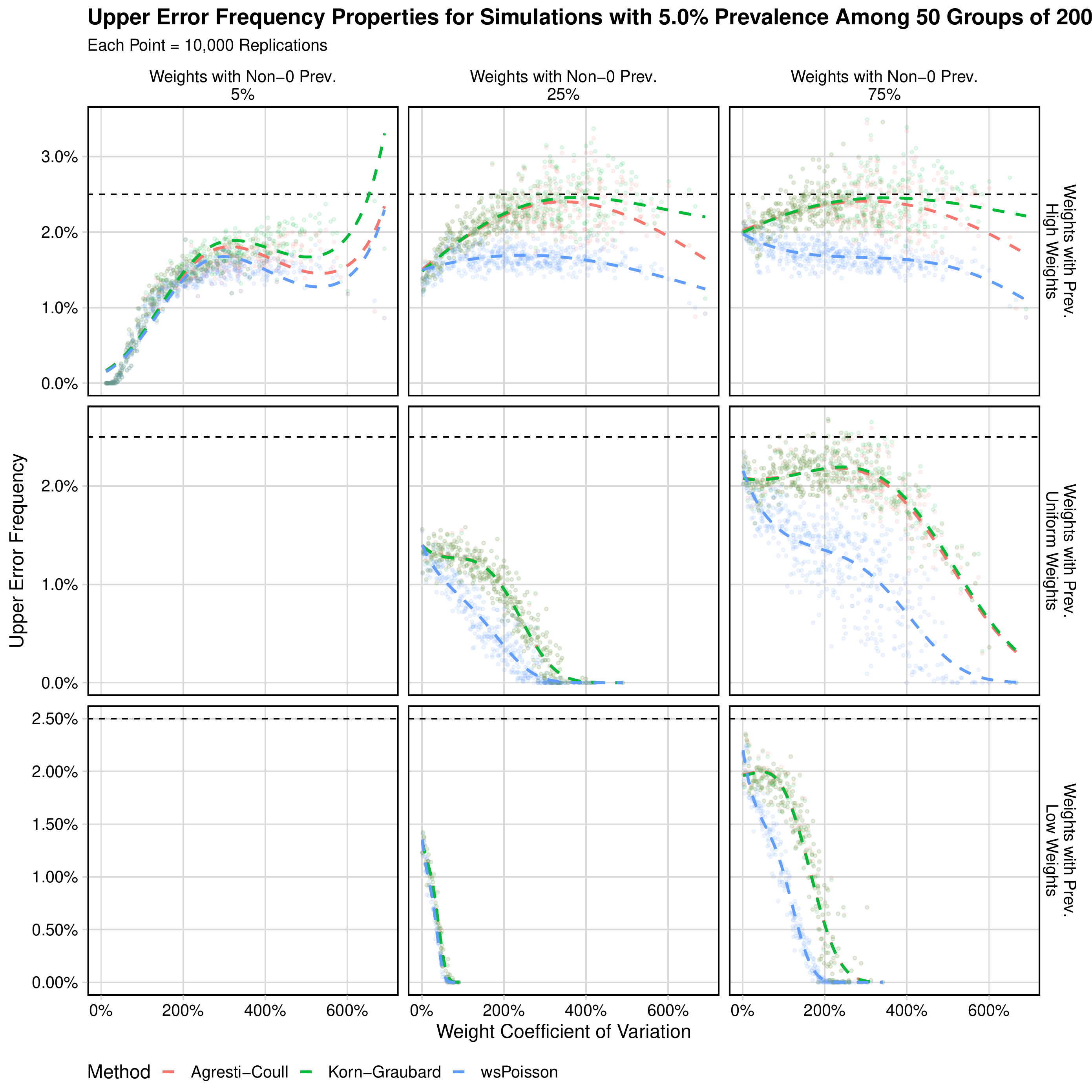}
\caption{Upper error properties for the wsPoisson model and two standard methods, the Dean-Pagano modification of the Agresti-Coull method and of the Korn-Graubard method.
Each point represents 10,000 simulations of datasets from a population with 5\% Prevalence where 50 groups of 200 people are sampled.
The horizontal dashed line indicates the nominal upper error rate, 2.5\%.
Colored dashed lines are estimates from a logistic regression model using quadratic splines.}
\label{fig:perfect_upper_error_frequency_50_groups_0_05_prev}
\end{figure}

\begin{figure}
\centering
\includegraphics[width=0.8\textwidth]{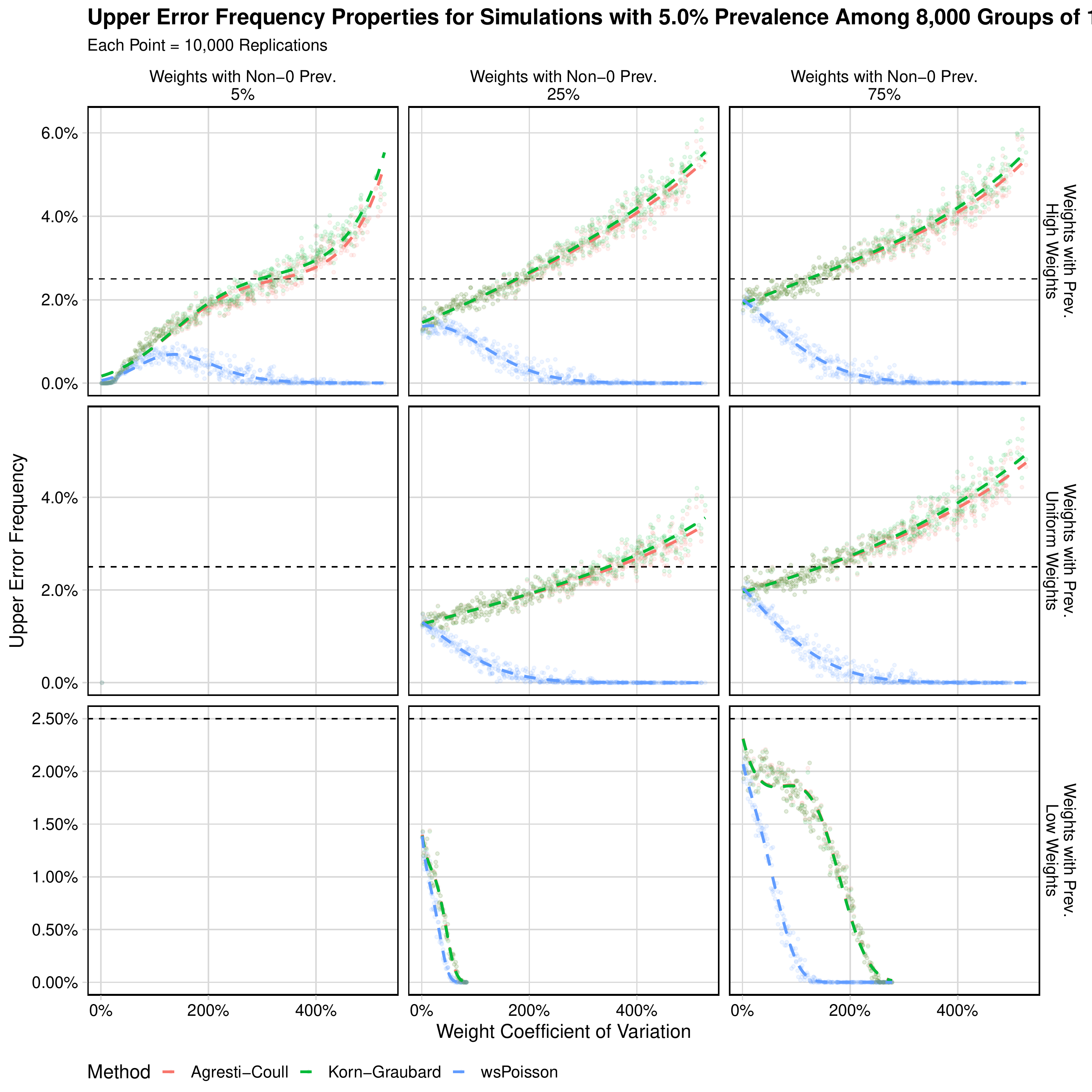}
\caption{Upper error properties for the wsPoisson model and two standard methods, the Dean-Pagano modification of the Agresti-Coull method and of the Korn-Graubard method.
Each point represents 10,000 simulations of datasets from a population with 5\% Prevalence where 8000 individuals are sampled.
The horizontal dashed line indicates the nominal upper error rate, 2.5\%.
Colored dashed lines are estimates from a logistic regression model using quadratic splines.}
\label{fig:perfect_upper_error_frequency_8000_groups_0_05_prev}
\end{figure}

\begin{figure}
\centering
\includegraphics[width=0.8\textwidth]{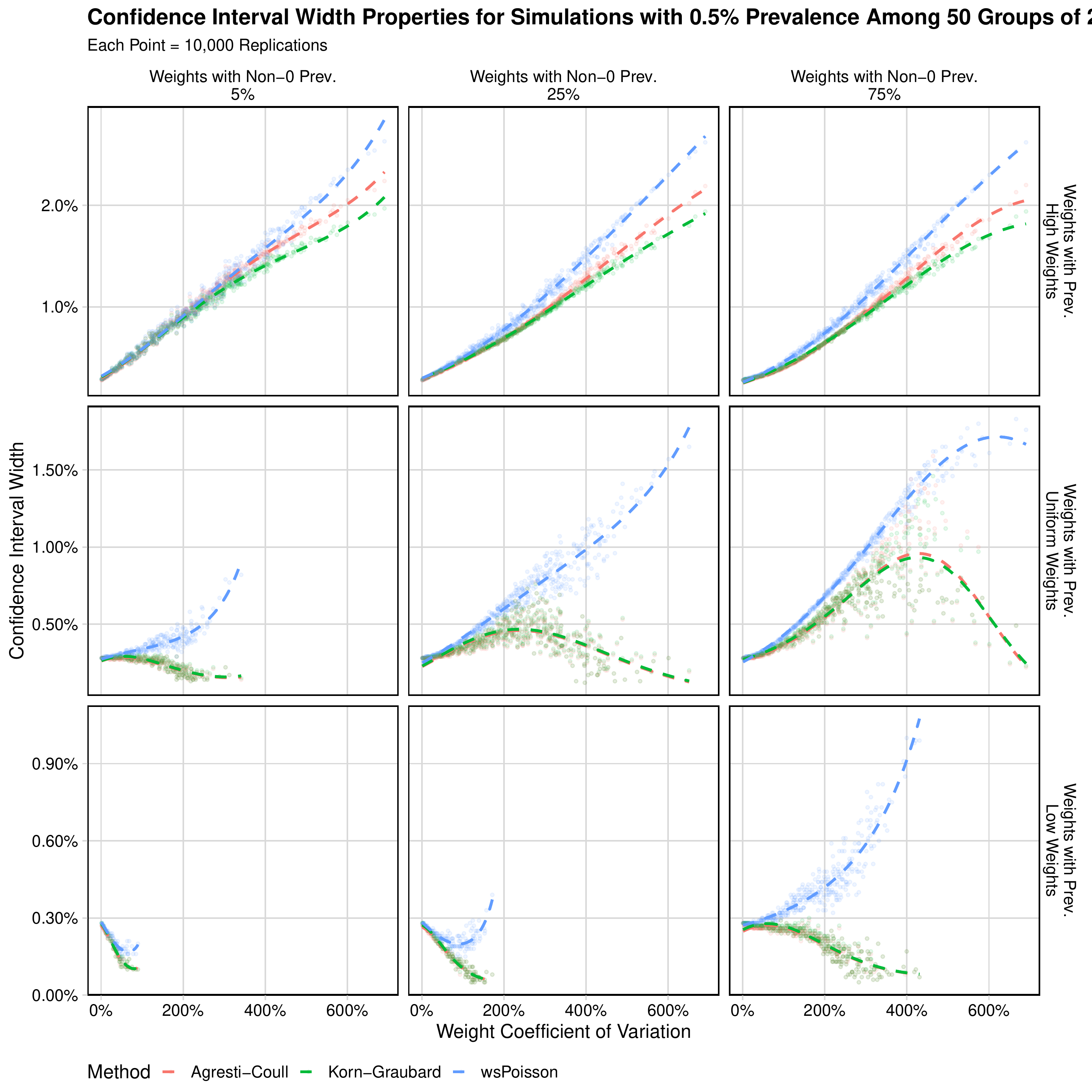}
\caption{Confidence interval width properties for the wsPoisson model and two standard methods, the Dean-Pagano modification of the Agresti-Coull method and of the Korn-Graubard method.
Each point represents 10,000 simulations of datasets from a population with 0.5\% Prevalence where 50 groups of 200 people are sampled.
Colored dashed lines are estimates from a logistic regression model using quadratic splines.}
\label{fig:perfect_confidence_interval_width_50_groups_0_005_prev}
\end{figure}

\begin{figure}
\centering
\includegraphics[width=0.8\textwidth]{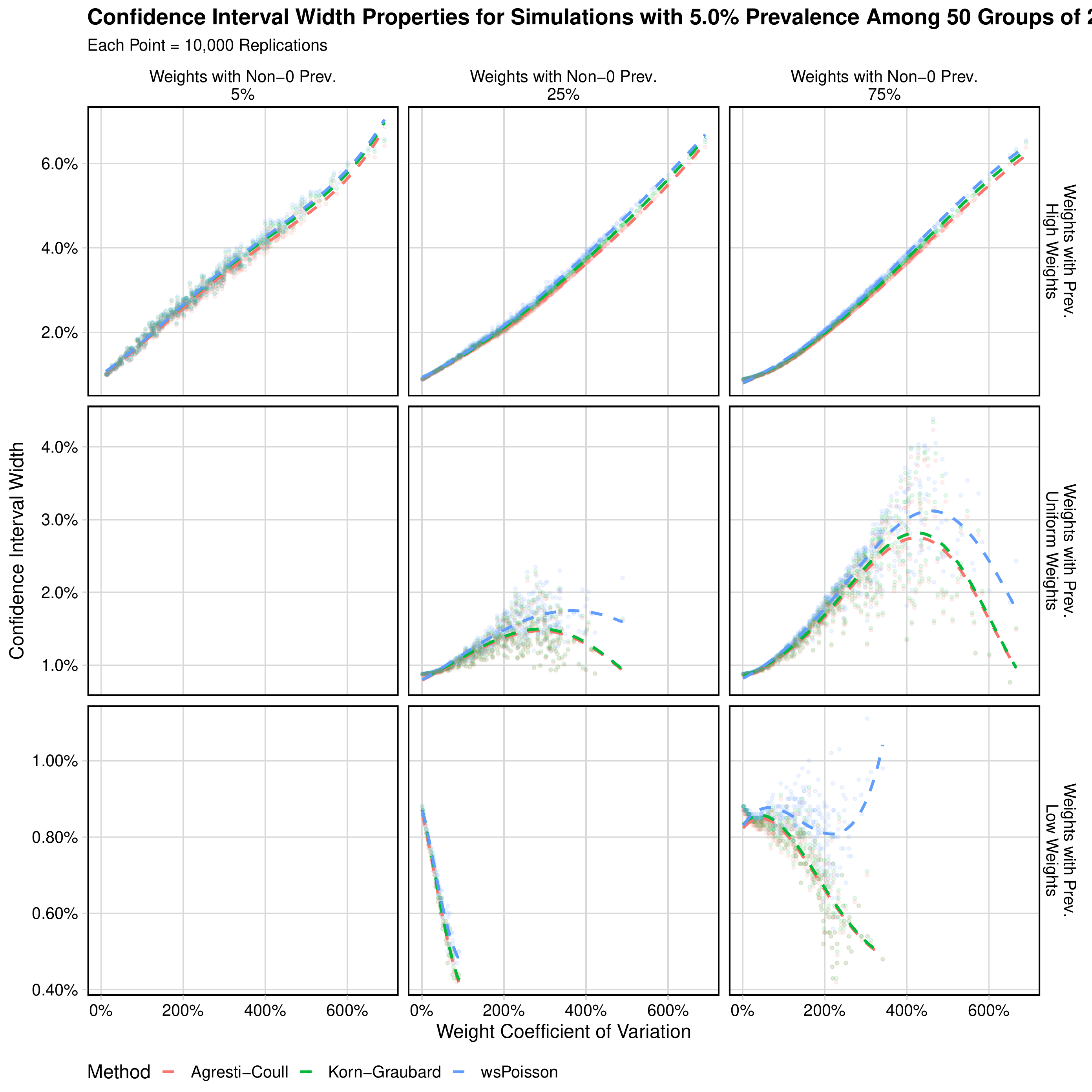}
\caption{Confidence interval width properties for the wsPoisson model and two standard methods, the Dean-Pagano modification of the Agresti-Coull method and of the Korn-Graubard method.
Each point represents 10,000 simulations of datasets from a population with 5\% Prevalence where 50 groups of 200 people are sampled.
Colored dashed lines are estimates from a logistic regression model using quadratic splines.}
\label{fig:perfect_confidence_interval_width_50_groups_0_05_prev}
\end{figure}

\begin{figure}
\centering
\includegraphics[width=0.8\textwidth]{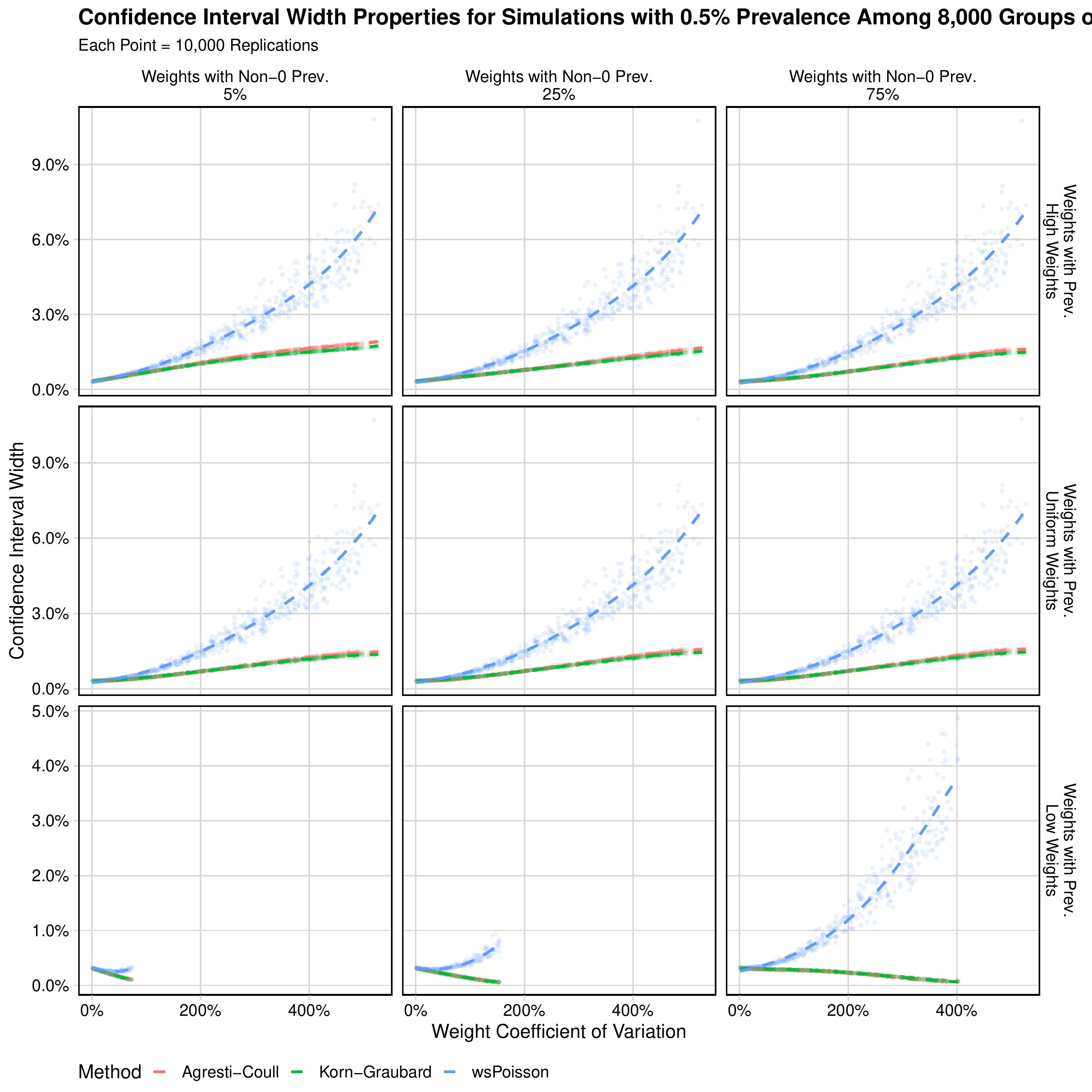}
\caption{Confidence interval width properties for the wsPoisson model and two standard methods, the Dean-Pagano modification of the Agresti-Coull method and of the Korn-Graubard method.
Each point represents 10,000 simulations of datasets from a population with 0.5\% Prevalence where 8000 individuals are sampled.
Colored dashed lines are estimates from a logistic regression model using quadratic splines.}
\label{fig:perfect_confidence_interval_width_8000_groups_0_005_prev}
\end{figure}

\begin{figure}
\centering
\includegraphics[width=0.8\textwidth]{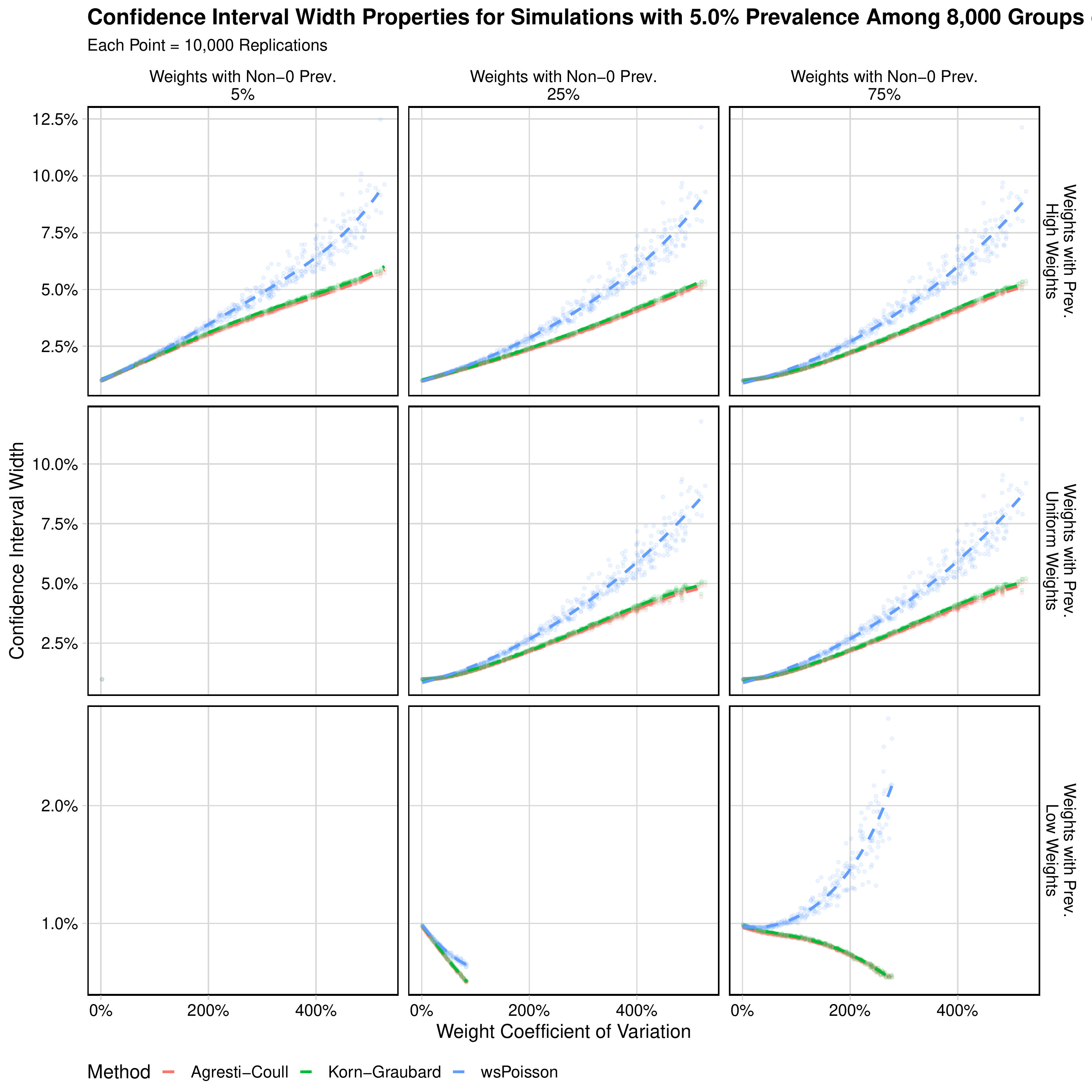}
\caption{Confidence interval width properties for the wsPoisson model and two standard methods, the Dean-Pagano modification of the Agresti-Coull method and of the Korn-Graubard method.
Each point represents 10,000 simulations of datasets from a population with 5\% Prevalence where 8000 individuals are sampled.
Colored dashed lines are estimates from a logistic regression model using quadratic splines.}
\label{fig:perfect_confidence_interval_width_8000_groups_0_05_prev}
\end{figure}


\begin{figure}
\centering
\includegraphics[width=0.8\textwidth]{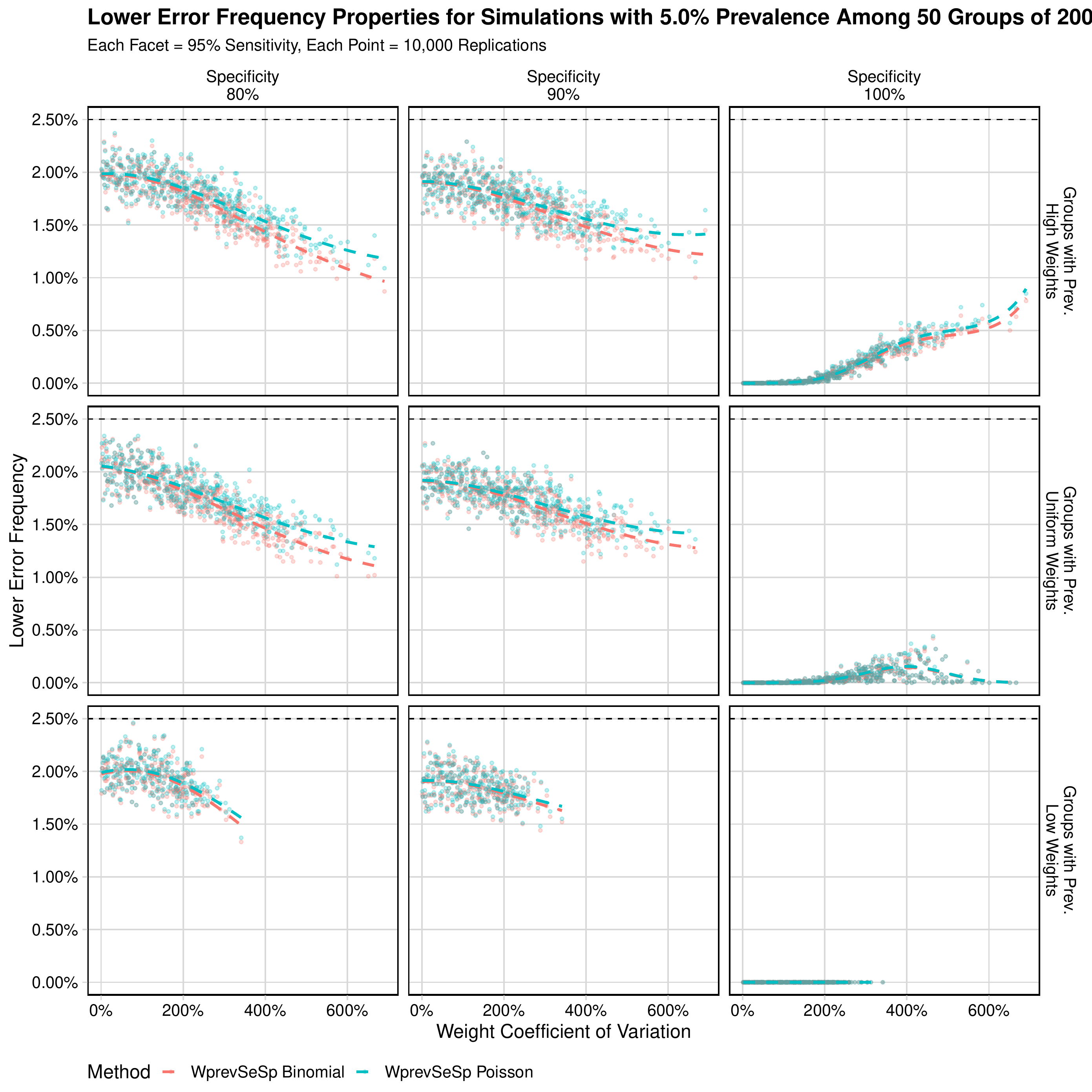}
\caption{Lower error properties for the confidence interval procedures, WprevSeSp Binomial and WprevSeSp Poisson.
Each point represents 10,000 simulations of datasets from a population with 0.5\% Prevalence where 50 groups of 200 people are sampled.
Each datasets also includes simulated results of tests to evaluate the sensitivity and specificity of the assay performed on 60 and 300 individuals, respectively.
The horizontal dashed line indicates the nominal lower error rate, 2.5\%.
Colored dashed lines are estimates from a logistic regression model using quadratic splines. If the WprevSeSp Binomial line is not visible, then it is covered by the WprevSeSp Poisson line.}
\label{fig:imperfect_lower_error_frequency_50_groups_0_05_prev}
\end{figure}

\begin{figure}
\centering
\includegraphics[width=0.8\textwidth]{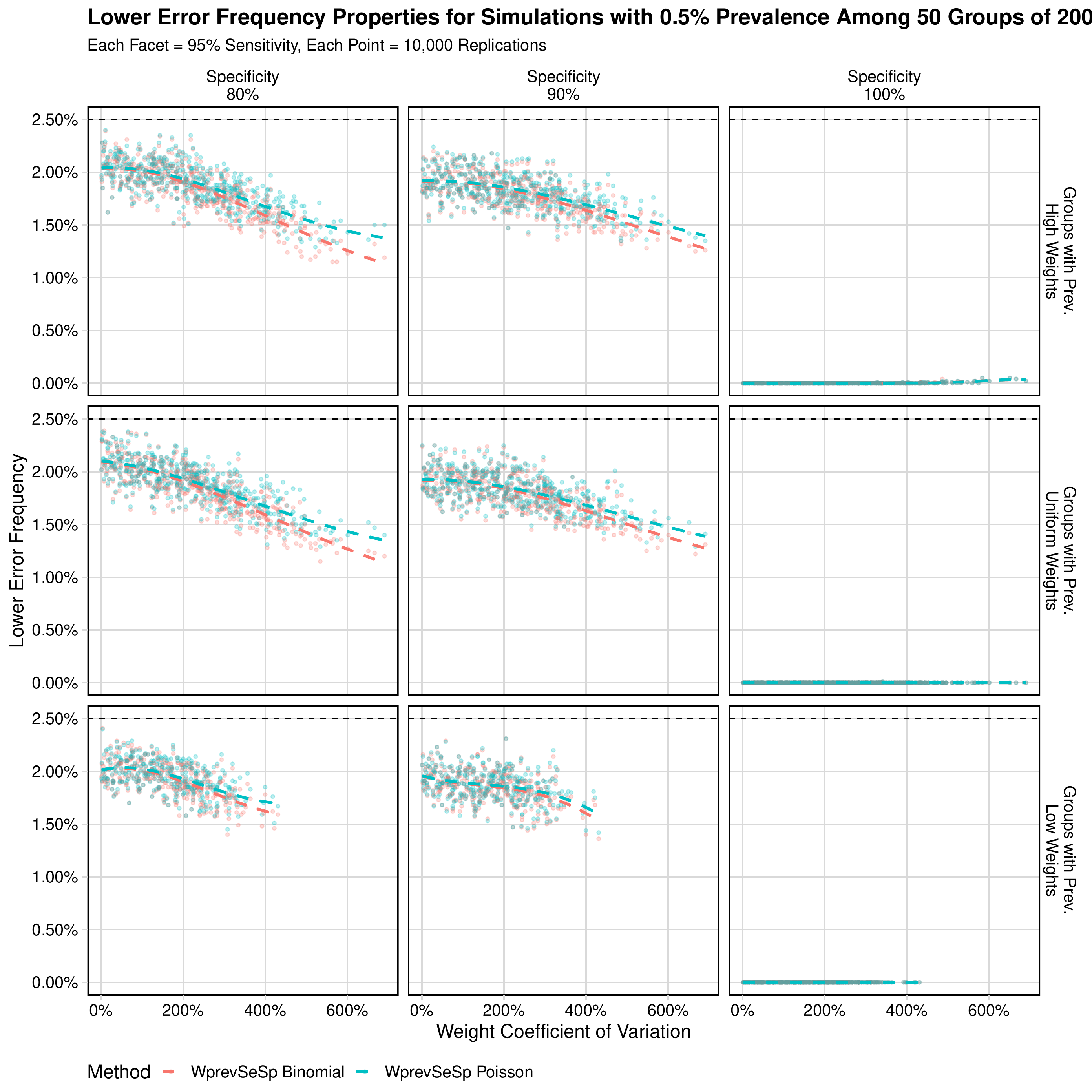}
\caption{Lower error properties for the confidence interval procedures, WprevSeSp Binomial and WprevSeSp Poisson.
Each point represents 10,000 simulations of datasets from a population with 0.5\% Prevalence where 8000 individuals are sampled.
Each datasets also includes simulated results of tests to evaluate the sensitivity and specificity of the assay performed on 60 and 300 individuals, respectively.
The horizontal dashed line indicates the nominal lower error rate, 2.5\%.
Colored dashed lines are estimates from a logistic regression model using quadratic splines.}
\label{fig:imperfect_lower_error_frequency_50_groups_0_005_prev}
\end{figure}

\begin{figure}
\centering
\includegraphics[width=0.8\textwidth]{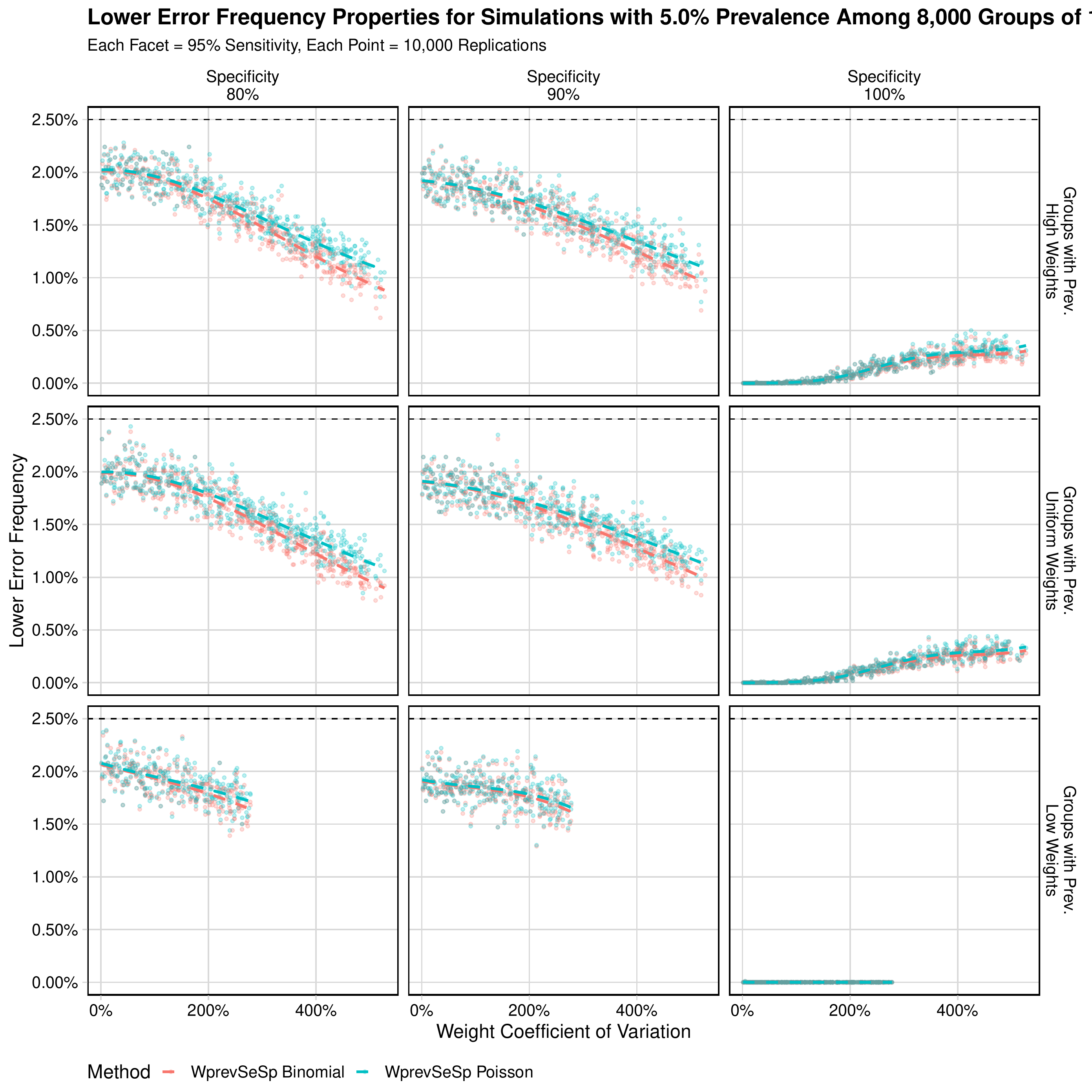}
\caption{Lower error properties for the confidence interval procedures, WprevSeSp Binomial and WprevSeSp Poisson.
Each point represents 10,000 simulations of datasets from a population with 5\% Prevalence where 50 groups of 200 people are sampled.
Each datasets also includes simulated results of tests to evaluate the sensitivity and specificity of the assay performed on 60 and 300 individuals, respectively.
The horizontal dashed line indicates the nominal lower error rate, 2.5\%.
Colored dashed lines are estimates from a logistic regression model using quadratic splines.}
\label{fig:imperfect_lower_error_frequency_8000_groups_0_05_prev}
\end{figure}

\begin{figure}
\centering
\includegraphics[width=0.8\textwidth]{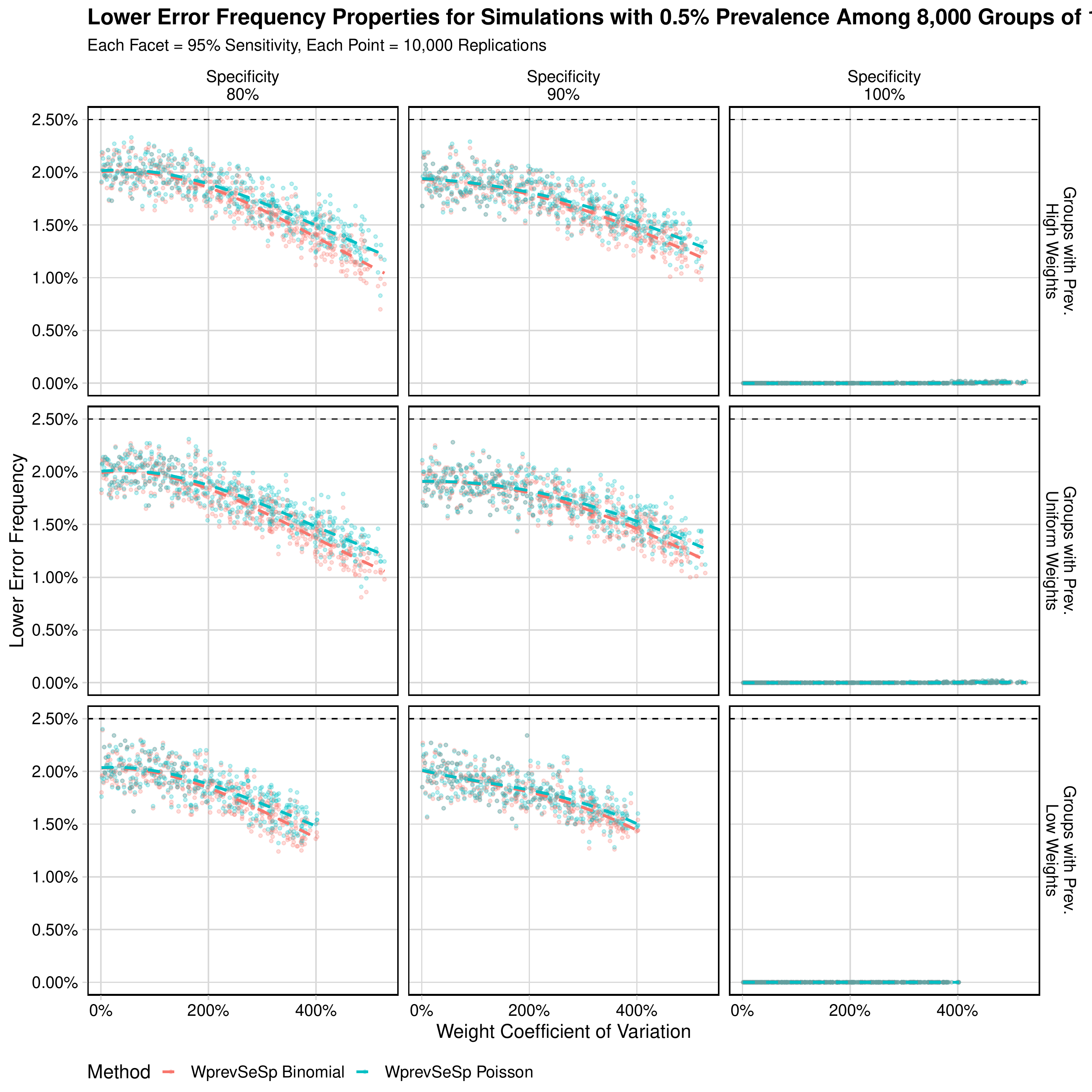}
\caption{Lower error properties for the confidence interval procedures, WprevSeSp Binomial and WprevSeSp Poisson.
Each point represents 10,000 simulations of datasets from a population with 5\% Prevalence where 8000 individuals are sampled.
Each datasets also includes simulated results of tests to evaluate the sensitivity and specificity of the assay performed on 60 and 300 individuals, respectively.
The horizontal dashed line indicates the nominal lower error rate, 2.5\%.
Colored dashed lines are estimates from a logistic regression model using quadratic splines.}
\label{fig:imperfect_lower_error_frequency_8000_groups_0_005_prev}
\end{figure}

\begin{figure}
\centering
\includegraphics[width=0.8\textwidth]{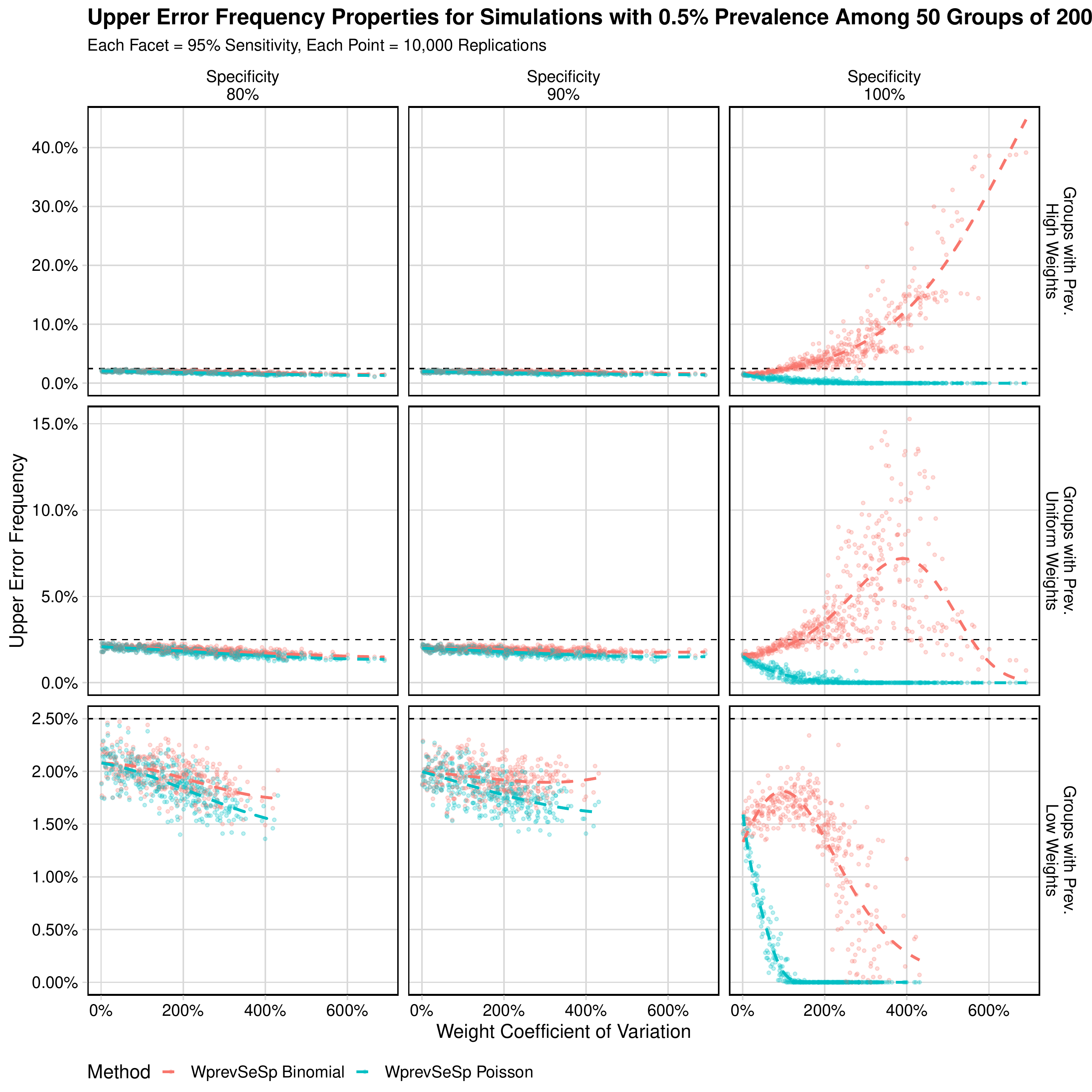}
\caption{Upper error properties for the confidence interval procedures, WprevSeSp Binomial and WprevSeSp Poisson.
Each point represents 10,000 simulations of datasets from a population with 0.5\% Prevalence where 50 groups of 200 people are sampled.
Each datasets also includes simulated results of tests to evaluate the sensitivity and specificity of the assay performed on 60 and 300 individuals, respectively.
The horizontal dashed line indicates the nominal upper error rate, 2.5\%.
Colored dashed lines are estimates from a logistic regression model using quadratic splines.}
\label{fig:imperfect_upper_error_frequency_50_groups_0_005_prev}
\end{figure}

\begin{figure}
\centering
\includegraphics[width=0.8\textwidth]{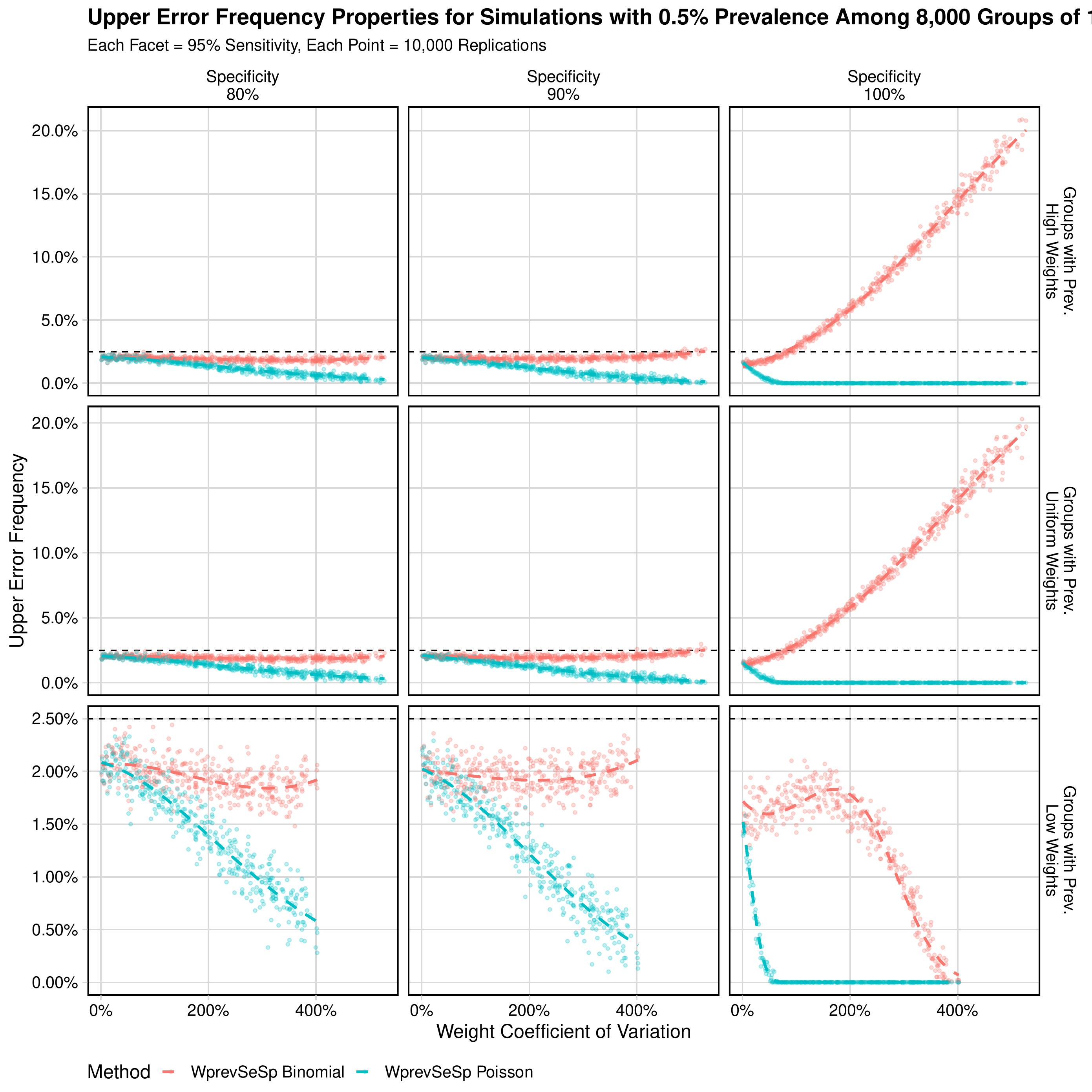}
\caption{Upper error properties for the confidence interval procedures, WprevSeSp Binomial and WprevSeSp Poisson.
Each point represents 10,000 simulations of datasets from a population with 0.5\% Prevalence where 8000 individuals are sampled.
Each datasets also includes simulated results of tests to evaluate the sensitivity and specificity of the assay performed on 60 and 300 individuals, respectively.
The horizontal dashed line indicates the nominal upper error rate, 2.5\%.
Colored dashed lines are estimates from a logistic regression model using quadratic splines.}
\label{fig:imperfect_upper_error_frequency_8000_groups_0_005_prev}
\end{figure}

\begin{figure}
\centering
\includegraphics[width=0.8\textwidth]{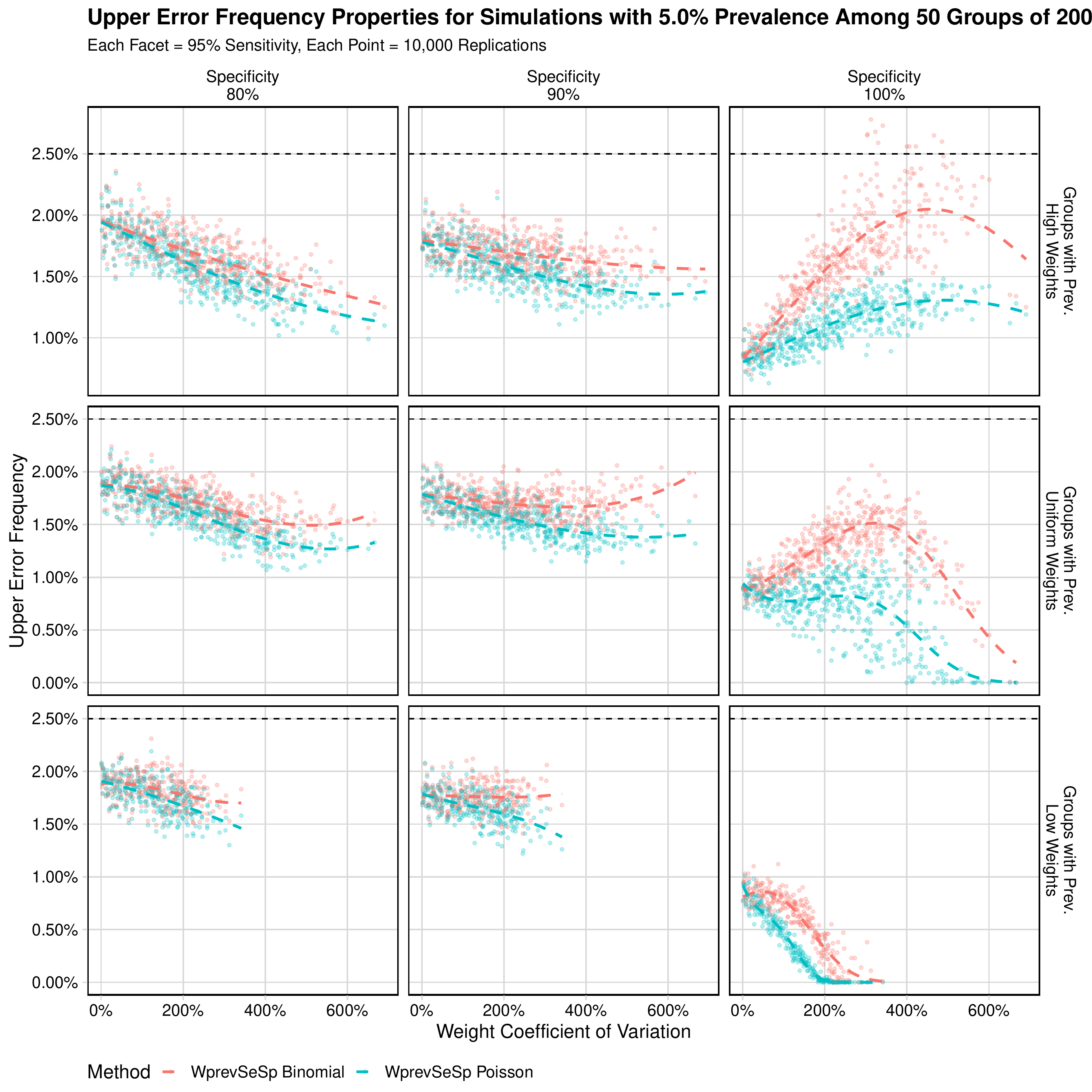}
\caption{Upper error properties for the confidence interval procedures, WprevSeSp Binomial and WprevSeSp Poisson.
Each point represents 10,000 simulations of datasets from a population with 5\% Prevalence where 50 groups of 200 people are sampled.
Each datasets also includes simulated results of tests to evaluate the sensitivity and specificity of the assay performed on 60 and 300 individuals, respectively.
The horizontal dashed line indicates the nominal upper error rate, 2.5\%.
Colored dashed lines are estimates from a logistic regression model using quadratic splines.}
\label{fig:imperfect_upper_error_frequency_50_groups_0_05_prev}
\end{figure}

\begin{figure}
\centering
\includegraphics[width=0.8\textwidth]{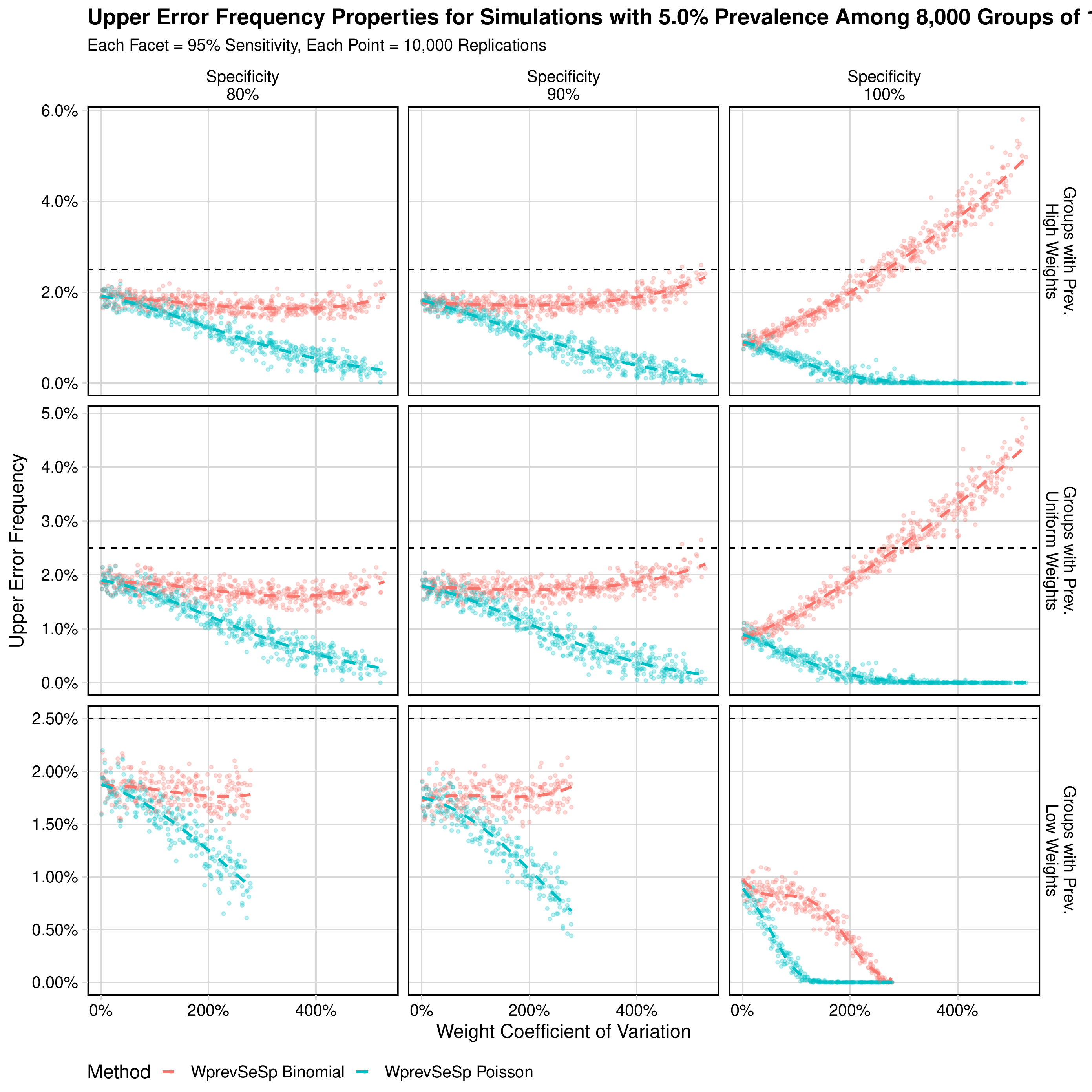}
\caption{Upper error properties for the confidence interval procedures, WprevSeSp Binomial and WprevSeSp Poisson.
Each point represents 10,000 simulations of datasets from a population with 5\% Prevalence where 8000 individuals are sampled.
Each datasets also includes simulated results of tests to evaluate the sensitivity and specificity of the assay performed on 60 and 300 individuals, respectively.
The horizontal dashed line indicates the nominal upper error rate, 2.5\%.
Colored dashed lines are estimates from a logistic regression model using quadratic splines.}
\label{fig:imperfect_upper_error_frequency_8000_groups_0_05_prev}
\end{figure}

\begin{figure}
\centering
\includegraphics[width=0.8\textwidth]{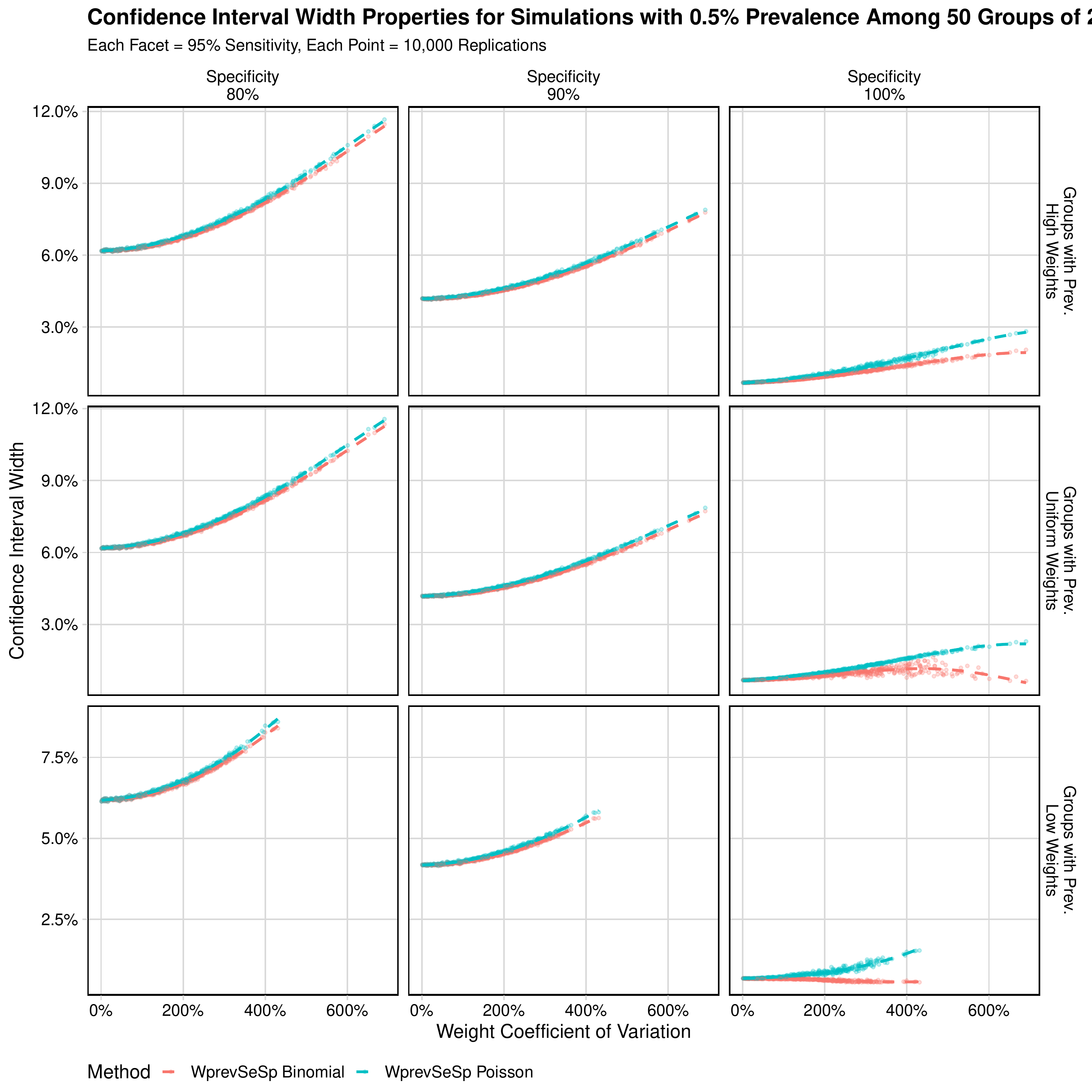}
\caption{Confidence interval width properties for the confidence interval procedures, WprevSeSp Binomial and WprevSeSp Poisson.
Each point represents 10,000 simulations of datasets from a population with 0.5\% Prevalence where 50 groups of 200 people are sampled.
Each datasets also includes simulated results of tests to evaluate the sensitivity and specificity of the assay performed on 60 and 300 individuals, respectively.
Colored dashed lines are estimates from a logistic regression model using quadratic splines.}
\label{fig:imperfect_confidence_interval_width_50_groups_0_005_prev}
\end{figure}

\begin{figure}
\centering
\includegraphics[width=0.8\textwidth]{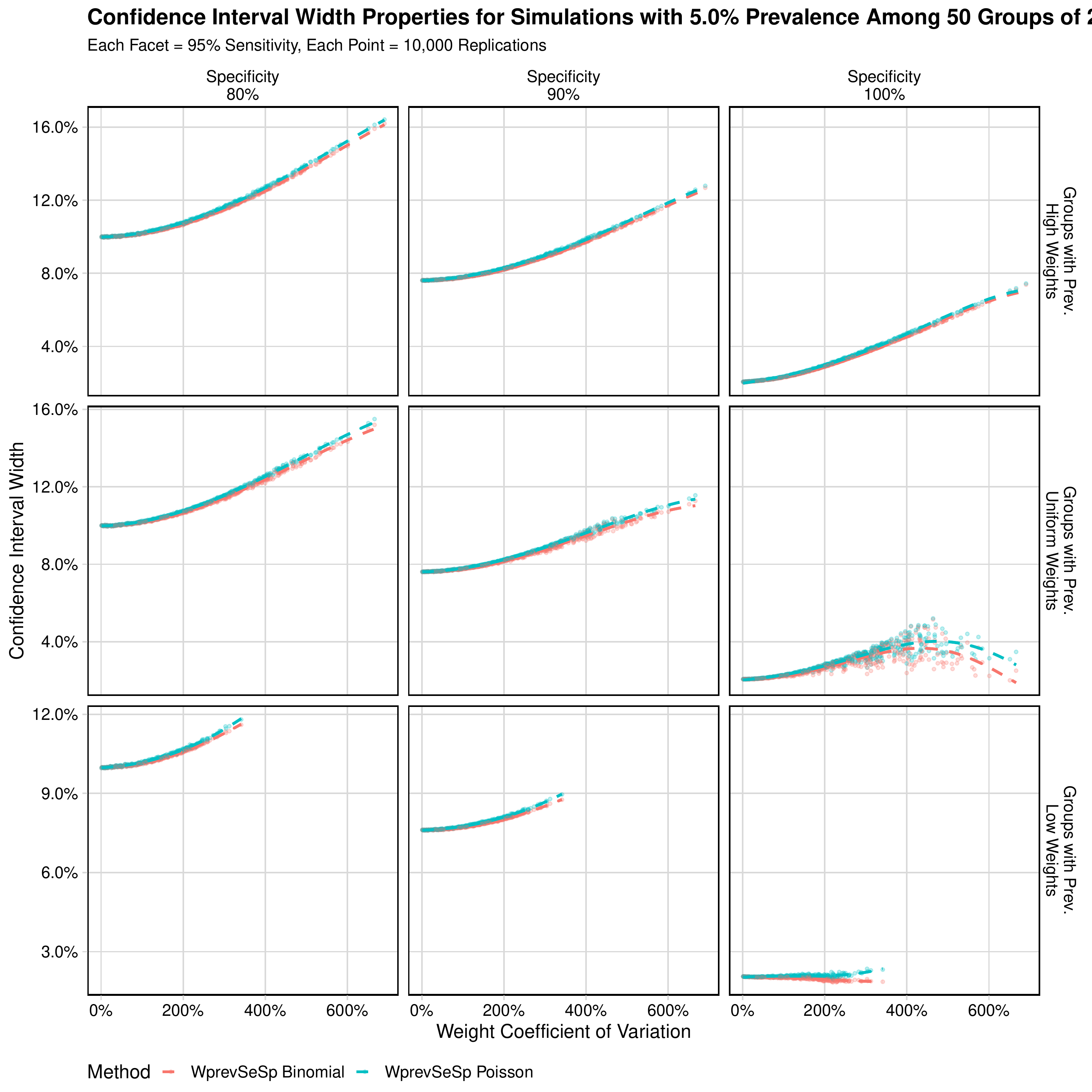}
\caption{Confidence interval width properties for the confidence interval procedures, WprevSeSp Binomial and WprevSeSp Poisson.
Each point represents 10,000 simulations of datasets from a population with 5\% Prevalence where 50 groups of 200 people are sampled.
Each datasets also includes simulated results of tests to evaluate the sensitivity and specificity of the assay performed on 60 and 300 individuals, respectively.
Colored dashed lines are estimates from a logistic regression model using quadratic splines.}
\label{fig:imperfect_confidence_interval_width_50_groups_0_05_prev}
\end{figure}

\begin{figure}
\centering
\includegraphics[width=0.8\textwidth]{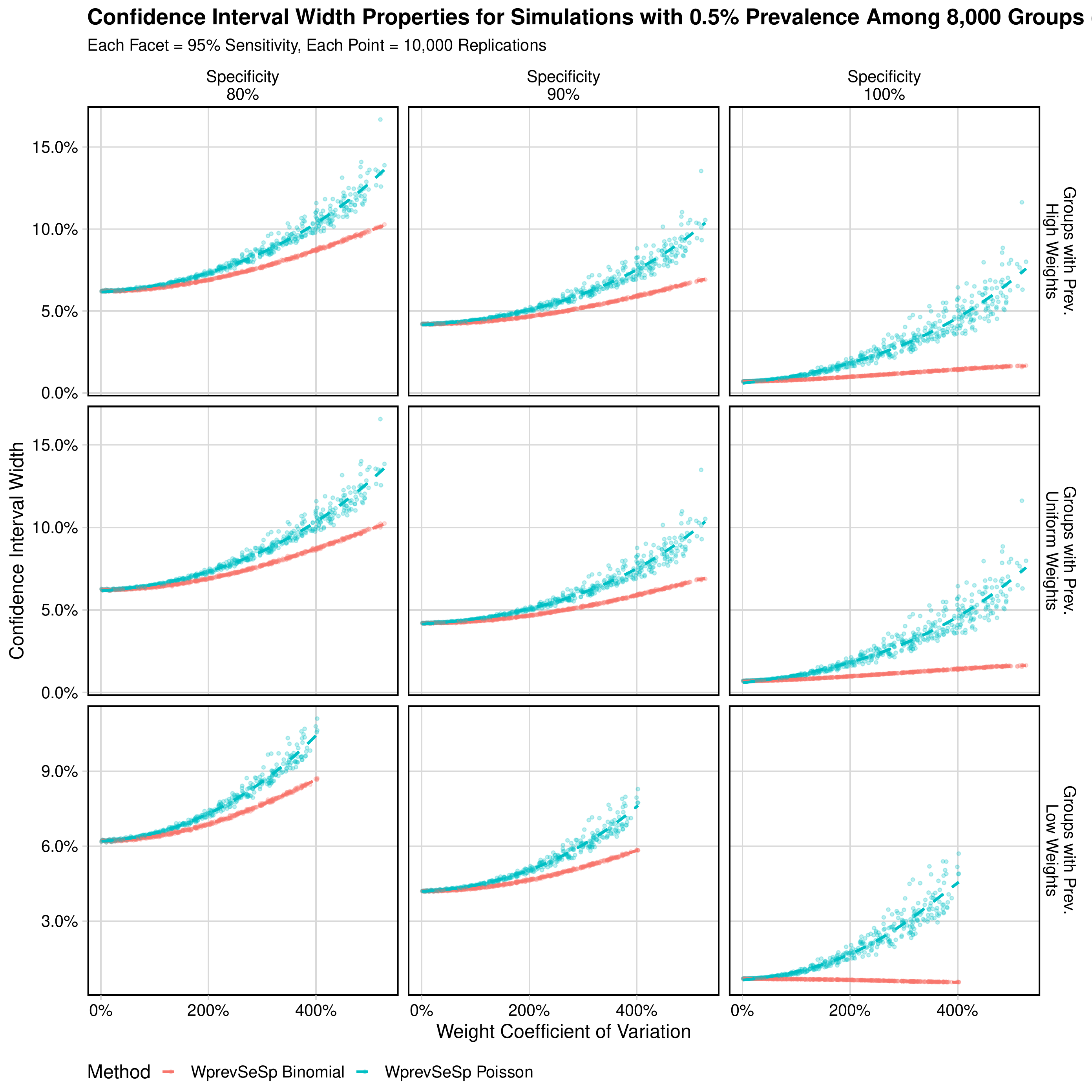}
\caption{Confidence interval width properties for the confidence interval procedures, WprevSeSp Binomial and WprevSeSp Poisson.
Each point represents 10,000 simulations of datasets from a population with 0.5\% Prevalence where 8000 individuals are sampled.
Each datasets also includes simulated results of tests to evaluate the sensitivity and specificity of the assay performed on 60 and 300 individuals, respectively.
Colored dashed lines are estimates from a logistic regression model using quadratic splines.}
\label{fig:imperfect_confidence_interval_width_8000_groups_0_005_prev}
\end{figure}

\begin{figure}
\centering
\includegraphics[width=0.8\textwidth]{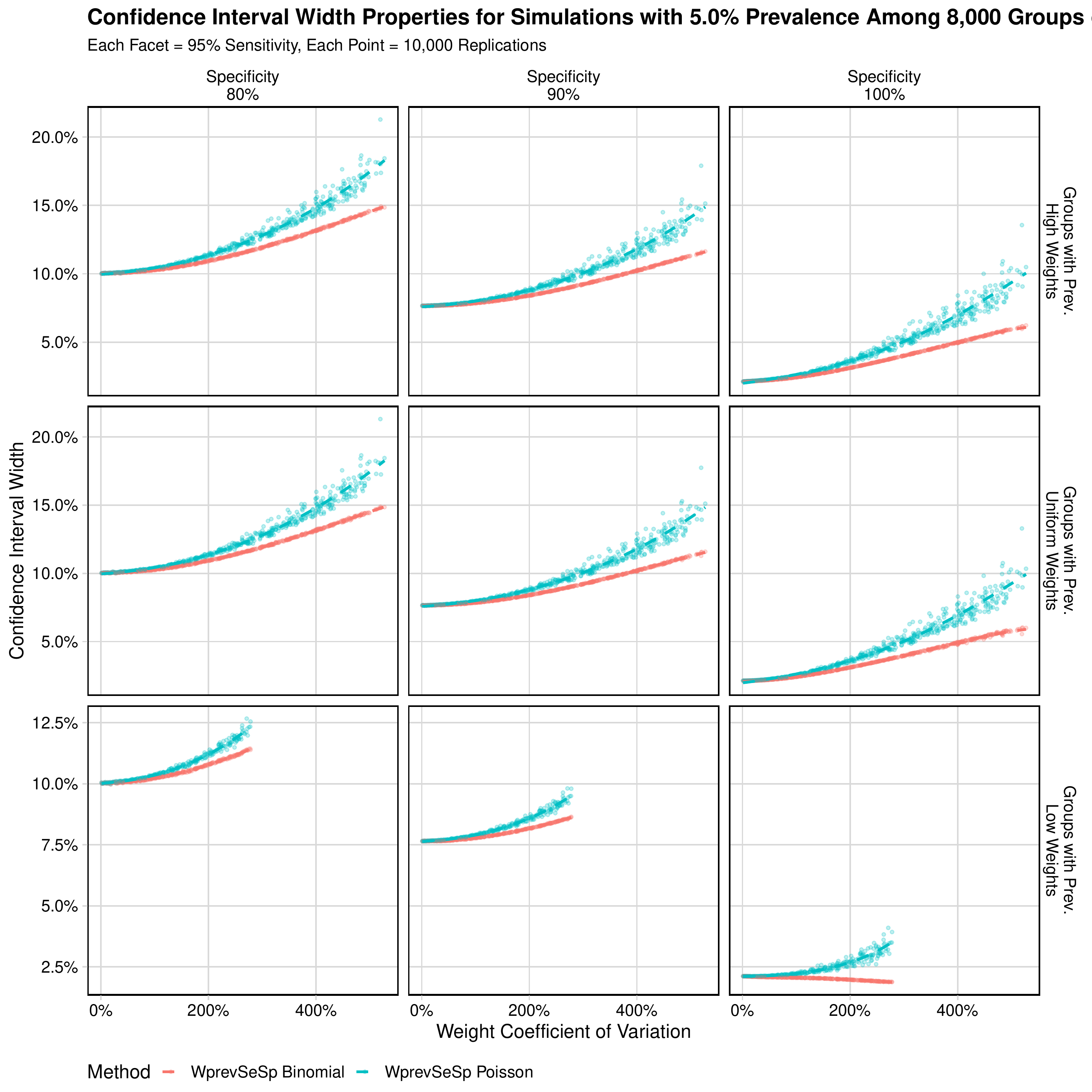}
\caption{Confidence interval width properties for the confidence interval procedures, WprevSeSp Binomial and WprevSeSp Poisson.
Each point represents 10,000 simulations of datasets from a population with 5\% Prevalence where 8000 individuals are sampled.
Each datasets also includes simulated results of tests to evaluate the sensitivity and specificity of the assay performed on 60 and 300 individuals, respectively.
Colored dashed lines are estimates from a logistic regression model using quadratic splines.}
\label{fig:imperfect_confidence_interval_width_8000_groups_0_05_prev}
\end{figure}

\end{document}